\documentclass[%
superscriptaddress,
 amsmath,amssymb,
prl,
twocolumn,
]{revtex4-2}
\usepackage[colorlinks]{hyperref}
\usepackage{lmodern}
\usepackage{amsfonts, amssymb, amsmath}
\usepackage{graphicx}
\usepackage{color}
\usepackage{lineno}
\usepackage[usenames,dvipsnames]{xcolor}
\usepackage[utf8]{inputenc}
\usepackage{mathtools}
\usepackage{esint}
\usepackage{bm}
\usepackage{natbib}
\usepackage{stackrel}
\usepackage{subcaption}
\usepackage{placeins}
\usepackage{amsmath}
\usepackage{soul}
\usepackage{latexsym}
\usepackage{graphicx,mathtools}
\usepackage{calligra}
\usepackage{upgreek}
\usepackage{color}
\usepackage{inputenc,caption}
\usepackage{stackrel}
\usepackage{subcaption}
\usepackage{placeins}
\usepackage{amsmath}
\usepackage{soul}
\usepackage{latexsym}

\definecolor{mygreen}{rgb}{0,0.5,0}
\definecolor{mybrown}{rgb}{0.65,0.16,0.16}
\definecolor{matlabgreen}{rgb}{0.466,0.674,0.188}
\definecolor{matcolv}{rgb}{0.494,0.184,0.556}

\def\ur{\mathbf{\ell}}
\def\uu{\mathbf{u}}
\def\ux{\mathbf{x}}

\def\beq {\begin{equation}}
\def\eeq {\end{equation}}
\def\beqa {\begin{eqnarray}}
\def\eeqa {\end{eqnarray}}
\def \bnum {\begin{enumerate}}
\def \enum {\end{enumerate}}
\def\bi {\begin{itemize}}
\def\ei {\end{itemize}}
\def \bdes {\begin{description}}
\def \edes {\end{description}}

\def\la {\langle}
\def\ra {\rangle}

\def\mbf {\mathbf}

\def\rmd {{\rm d}}
\def\zith{\zeta^\theta_p}

\def\ux{\mbf x}
\def\ur{\boldsymbol{\ell}}

\def\zith{\zeta_3}
\def\zithnu{\zeta_3^\nu}

\def\bu{{\bf u}^\nu}

\def\Re{\mathsf{Re}}
\def\U {{u^\prime}}

\def\br{{\boldsymbol{\ell}}} 
\def\bu{{\bf u}} 
\def\bx{{\bf x}} 
\def\dstar{{D_{*}}}

\def\srv{S_{3}(\br)}
\def\sr{S_{3}(\ell)}

\def\srlong{S_{3,\parallel}(\ell)}
\def\elo{\mathsf{L}}

\def\drulong{\delta_\ell u_\parallel^\nu (\ux,t)}

\def\druv{\delta_{\boldsymbol{\ell}} {\uu}(\ux,t)}

\def\epsn{\varepsilon}

\def\U {u^\prime}

\def\du{ D[\bu]}

\def\re{\mathsf{Re}}

\def\lbox{\mathsf{L}_\mathsf{box}}
\def\L{\mathsf{L}}
\def\U{\mathsf{U}}

\newcommand{\be}{\begin{equation}}
\newcommand{\ee}{\end{equation}}

\begin{document}
\title{Whither the Zeroth Law of Turbulence?}
\author{Kartik P. Iyer}
\affiliation{Department of Physics and Department of Mechanical and Aerospace Engineering, Michigan Technological University, Houghton, MI 49931, USA}
\email{kiyer@mtu.edu}
\author{Theodore D. Drivas}
\affiliation{Mathematics Department, Stony Brook University, Stony Brook, NY 11794, USA}\email{tdrivas@math.stonybrook.edu}
\author{Gregory L. Eyink}
\affiliation{Department of Applied Mathematics \& Statistics and Department of Physics \& Astronomy, The Johns Hopkins University, Baltimore, MD 21218, USA}
\author{Katepalli R. Sreenivasan}
\affiliation{Department of Physics, and the Courant Institute of Mathematical Sciences, New York University, New York, NY 11201, USA}

\begin{abstract}
Experimental and numerical studies of incompressible turbulence suggest that the mean dissipation rate of kinetic energy remains {constant} as the Reynolds number tends to infinity (or the non-dimensional viscosity tends to zero). 
This anomalous behavior is central to many theories of high-Reynolds-number turbulence and for this reason 
has been termed the ``zeroth law". Here we report  
a sequence of direct numerical simulations of incompressible Navier-Stokes in a box with periodic boundary conditions, which indicate that the anomaly vanishes at a rate that 
agrees with the scaling of third-moment of absolute velocity increments.
Our results suggest that turbulence without boundaries may not develop strong enough singularities to sustain the zeroth law.
\end{abstract}

\maketitle

A central object in the study of incompressible fluid turbulence is the average dissipation rate of kinetic energy.  
For a velocity field $\bu$ obeying the incompressible Navier-Stokes equation with kinematic viscosity $\nu,$ the average dissipation (per unit mass) {may be} written as 
\be\label{dissipation}
\varepsilon[\bu]  =  \frac{1}{T}\int_0^T \!\!\! \rmd t\ \frac{1}{|\Omega|}\int_{\Omega}\! \rmd V \ \nu |\nabla \bu(\bx,t)|^2 ,
\ee
where the joint average is with respect to volume over the flow domain $\Omega$ and time over 
an interval $[0,T],$
although only one of these averages may be taken in practice. Because of the prefactor $\nu$ in the integrand, the mean dissipation 
normalized by a characteristic large-scale velocity $\U$ and a characteristic large length scale $\L$, 
\be\label{Ddef}
 D[\bu] = \frac{\varepsilon}{{\U}^3/\L},  
\ee
would na\"{\i}vely be expected to decay as $\Re^{-1}$ where $\Re = \U\L/\nu$ is the {corresponding} Reynolds number. 
This decay is indeed observed in laminar flows but experiments and simulations in turbulent flows suggest another behavior,
\be\label{zerothlaw}
 D[\bu] \to D_*>0 \ \ \ \ \text{as} \ \ \ \ \Re\to \infty.
\ee
In other words, energy dissipation persists even as the non-dimensional viscosity $\nu/\U \L$ vanishes. 
Evidence for this remarkable behavior is varied: experiments in grid turbulence
\cite{sreenivasan1984scaling,pearson2002measurements}, wakes behind bluff bodies and jets 
\cite{sreenivasan1995energy,pearson2002measurements},
 and the existence of a limiting aerodynamic drag coefficient $C_D$ 
for a moving body when $\Re \gg 1$ (see \cite{frisch1995turbulence}, section $5.2$). 
There is evidence for \eqref{zerothlaw} in internal flows as well, but intriguingly only when the solid walls 
are hydraulically rough, as in pipe flow with sand-grain roughness \cite{nikuradse1933laws} or in Taylor-Couette cells with side 
ridges or von K\'arm\'an flows with bladed stirrrers \cite{cadot1997energy}. Instead, the dimensionless energy dissipation rate 
is seen to decay to zero, although far more slowly than $1/\Re,$ in smooth-wall pipes \cite{mckeon2004friction} or in Taylor-Couette 
and von K\'arm\'an flows with smooth walls and stirrers \cite{cadot1997energy}. Other crucial evidence to complement 
this picture comes from direct numerical simulations (DNS) of homogeneous and 
isotropic turbulence (HIT) in a periodic box with smooth, large-scale body forces \cite{sreenivasan1998update,kaneda2003energy}. We re-examine this issue here because of its central importance to turbulence theory.

{\vspace{0.2cm}
The result \eqref{zerothlaw} was anticipated by G.~I. Taylor \cite{taylor1917observations} who later \cite{taylor1935statistical}
proposed a phenomenological basis for it. Relation \eqref{zerothlaw} was also assumed by Kolmogorov \cite{kolmogorov1941local,kolmogorov1941dissipation}. Because this result is fundamental to the theory of high-Reynolds turbulence, it is often called the ``zeroth law"~\footnote{To the best of our knowledge, this name was coined by one of us (KRS)
{sometime in the 1980's.}}. Another term
``dissipative anomaly'' arises because, the fluid motion without viscosity is governed by incompressible Euler equations, whose smooth solutions 
do not dissipate kinetic energy. To resolve this apparent contradiction,  Onsager \cite{onsager1949statistical} argued 
that energy dissipation in three-dimensional (3D) turbulence may not vanish in the zero-viscosity limit because of the lack of smoothness of the limiting velocity field.  
In particular, Onsager envisioned that the inviscid limits of Navier-Stokes solutions are described by singular weak (distributional) solutions 
of the Euler equations. Moreover, he stated that these solutions could possess at most a third 
of a derivative in the sense that they cannot be spatially H\"{o}lder continuous $C^{h}$ with any $h>1/3,$ else they would necessarily 
conserve energy. Constantin, E \& Titi  \cite{constantin1994onsager}  proved a stronger result for Euler solutions. 
Simply put, suppose the velocity possesses finite third moment and the third-order absolute structure function for the velocity increment 
$\delta_\br\bu(\bx,t) = \bu(\bx+\br,t)-\bu(\bx,t)$ satisfies 
 \be\label{sfunction}
 S_3(\br) =   \frac{1}{T}\int_0^T \!\!\! \rmd t\ \frac{1}{\Omega}\int_{\Omega}\! \rmd V  \ |\delta_\br\bu(\bx,t)|^3 \leq C_T \left(\frac{\ell}{\L}\right)^{\zeta_3}
 \ee
for all length scales $\ell \equiv |\br| <\L$, with the constant $C_T$ 
{independent of the Reynolds number}, the validity of the zeroth law 
\eqref{zerothlaw} requires $\zith \leq 1$ \cite{constantin1994onsager,duchon2000inertial,drivas2019onsager}.  
\begin{figure}[h!]
\includegraphics[width=1\columnwidth]{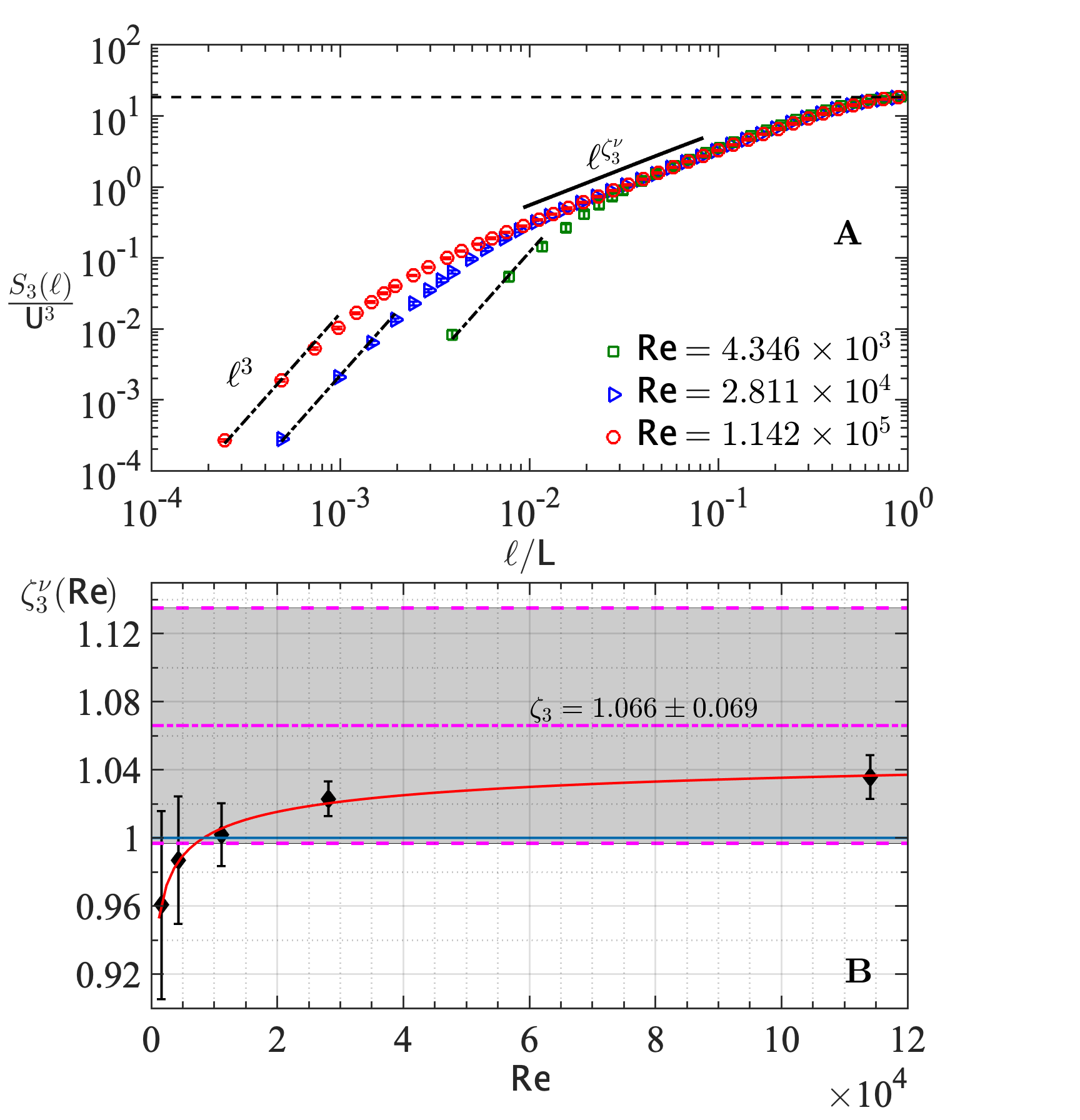}
\protect\caption{
Scaling of third-order absolute velocity structure function from the DNS of HIT in periodic cubes with edge-length $2\lbox$. ({\bf{A}}) Isotropic projections of $\srv$ normalized by cube of velocity scale, which is the root-mean-square velocity (see text), plotted against non-dimensional scale $\ell/\L$ with $\L = \lbox$ for three different Reynolds numbers {$\re = {\U \L}/{\nu}$}.
{Error bars corresponding to temporal variations of $\sr$ in statistically steady-state are smaller than symbol sizes at higher $\re$}. At smallest scales
$\ell/\eta \approx 1$, $\eta$ is the Kolmogorov length scale, $\sr$ displays cubic scaling indicated by
dash-dot lines, as expected. For the intermediate range of scales 
$\eta/\L \ll \ell/\L \ll 1,$
known as the inertial range (marked by filled symbols), a sub-cubic scaling range is seen, extending to increasingly smaller $\ell$ with increasing $\re$. Dashed horizontal line is the power-law prefactor $C_2$ in \eqref{sfbndns}. ({\bf{B}}) Exponents $\zithnu$ obtained by power-law fits to the inertial range are plotted as a function of $\re$. Error bars correspond to standard deviation of $\zithnu$ in statistically steady-state. Dotted line at unity is the naive Kolmogorov estimate \cite{kolmogorov1941local}. 
Solid curve is the least-squares fit $\zithnu = \zith + a {\Re}^b \ln {\Re}$  with $a=-0.284,$ $b = -0.406$ (see Supplemental Information \cite{SIpaper} for details); the asymptotic fit parameter $\zith$ along with its $95\%$ confidence interval (shaded region) are shown, with a root mean square error of $0.0032$. 
}
\label{fig1}
\end{figure}

{\vspace{0.2cm}
We elaborate here on this key result for our analysis. If there exist Reynolds number independent constants $C_1,$ $C_2$ (which may depend on averaging time $T$) such that
\be\label{sfbndns}
\frac{\langle |\bu|^3\rangle}{\U^3} \leq C_1(T), \qquad \frac{ S_3(\ell)}{\U^3}  \leq C_2(T)  \left(\frac{\ell}{\L}\right)^{\zeta_3}, 
\ee
 then a direct estimate for Navier-Stokes solutions \cite{drivas2019onsager} shows that the space-time averaged non-dimensional dissipation is bounded as
\be \label{expcond}
 D[\bu] \leq {\rm(const.)} {\Re}^{\frac{3(1-\zeta_3)}{3+ \zeta_3}}.
\ee
This bound implies that, if $\zith > 1$, then {$\du=O\left(\Re^{-\alpha}\right)$} 
with $\alpha \ge {{3(\zeta_3-1)}/{(3+ \zeta_3)}} > 0$; that is, (6) imposes a minimum decay rate on $\du$ if $\zith  >1$. 
Conversely, if {the zeroth law (3) holds with $\alpha=0,$ then the above bound 
requires $\zith \le 1$}. Since there exist constants $C_1$ \cite{KI23} and $C_2$ (see Figure \ref{fig1} below) that satisfy \eqref{sfbndns}, estimates of $\zith$ allow us to bound $\du$ using \eqref{expcond}.

 \begin{figure}[h!]
\includegraphics[width=1\columnwidth]{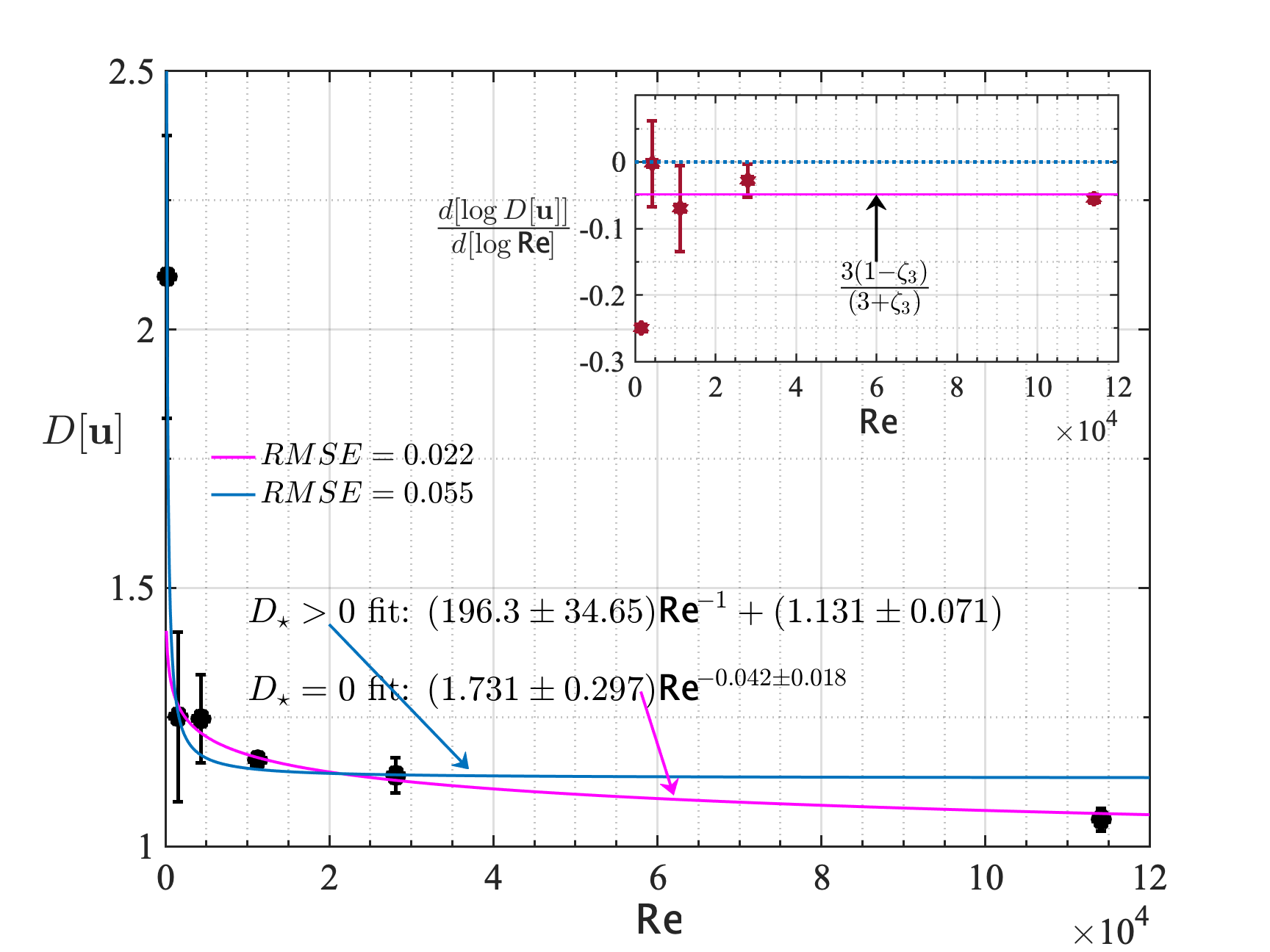}
\protect\caption{
Estimation of the asymptotic mean dissipation $\dstar$ based on the asymptotic third-order absolute velocity exponent $\zith$ from Figure \ref{fig1}. In the main panel the normalized mean dissipation $\du$ is plotted as a function of $\Re$. Error bars computed from standard deviation of $\du$ in statistically steady state are shown. Fits corresponding to vanishing asymptotic dissipation and non-zero asymptotic dissipation are given; also shown are the respective RMSE for both fits. Inset shows the logarithmic local slopes of $\du$ which give the dissipation exponent $\alpha$ as a function of $\Re$ without any curve fitting. Solid horizontal line is the decay exponent from bound \eqref{expcond} using $\zith = 1.066$ (from Figure \ref{fig1}), whereas dotted line at zero corresponds to $\zith \le 1$. 
}
\label{fig2}
\end{figure}

{\vspace{0.2cm}
For the Reynolds numbers considered here, Figure \ref{fig1} confirms the existence of the scaling law 
$\sr \sim \U^3 (\ell/\L)^{\zithnu}$ 
with viscosity-dependent (or $\Re$-dependent) exponents $\zithnu$ in a range of length scales known as the inertial range.
Here, $\U  = \sqrt{\langle |\uu|^2 \rangle/3}$, the root-mean-square (rms) velocity fluctuation, is chosen as the velocity scale $\U$ and half the linear size of the  computational domain, $\lbox$, is chosen as the relevant length scale $\L$.  We appreciate the difficulties in defining the most suitable length scale and discuss an alternate choice for $\L$ in the Supplemental Information \cite{SIpaper}.
The exponents $\zithnu$ may be inferred from plots of $\sr$ versus $\ell/\L$ using least-square power-law fits and/or logarithmic local slopes. We have followed standard procedures for the DNS data of HIT, which we describe more completely in the Supplemental Information \cite{SIpaper}. Results for $\zithnu$ plotted in Figure \ref{fig1} as a function of $\Re$ show a clear tendency for the exponents $\zithnu$ to rise past unity with increasing $\Re$, at least for the range shown here.
In the Supplementary Information we discuss the careful evaluation of $\zithnu$ including error bars \cite{SIpaper}. If the exponent trend found in Figure \ref{fig1} persists indefinitely, then our numerical results for HIT contradict the zeroth law requirement $\zith = \lim_{\Re \to 0} \zithnu \le 1$. The extrapolated value of $\zithnu$ using least squares fits will be taken below as $\zith$. Since we are estimating the fit parameter $\zith$ in the $\Re \to \infty$ limit -- a region in which the data become increasingly sparse -- the procedure has appreciable error bars as indicated. The asymptotic fit parameter $\zith$ is found to be larger than unity, however, our best estimates of $\zith$ do contain $\zith \le 1$ within error bars as indicated in Figure \ref{fig1}, which is consistent with the zeroth law. 
We should point out that past studies \cite{KRS94,Herweijer,Belin96,KRS98,Dhruva2000,GFN02,KRS22} have measured the longitudinal version of $\zithnu$ (see Supplementary Information \cite{SIpaper} for a discussion), but no consideration was given to its theoretical implications. 

\vspace{0.2cm}
Figure \ref{fig2} reports our fits to the normalized dissipation $\du$, and presents two possible fits consistent with $\zith > 1$ $(\dstar = 0)$  and $\zith \le 1$  $(\dstar > 0)$ \cite{DOERING_FOIAS_2002}. We note that the root mean square error (RMSE) for the  $\dstar = 0$ fits are smaller than those for the $\dstar > 0$ fit.
For the extrapolated estimate $\zith = 1.066$ from Figure \ref{fig1} (B), bound \eqref{expcond} implies that the dissipation decay exponent 
$\alpha$ must be at least $\frac{3(\zeta_3-1)}{3+ \zeta_3} =  0.0487$. We see that the power-law fit corresponding to $\dstar = 0$ yields an exponent remarkably close to this lower bound. This agreement between the dissipation decay exponent and the scaling bound \eqref{expcond} with increasing $\Re$ is further demonstrated in the inset which plots the dissipation local slopes $\alpha(\re)$. Local slopes are devoid of any curve fitting assumptions and thus offer a fit-free assessment of the scaling. With increasing $\Re$ the dissipation exponent $\alpha(\re)$ decreases towards the scaling bound \eqref{expcond}, away from anomalous case $\alpha = 0$.   

\begin{figure}[h!]
\includegraphics[width=1\columnwidth]{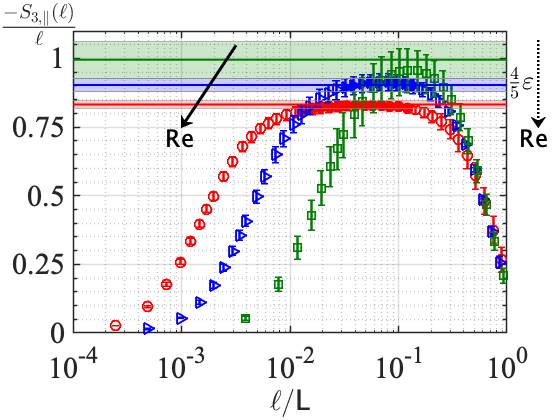} 
\protect\caption{Verification of the 4/5th law in 3D turbulence.
Symbols correspond to the third-order longitudinal structure function compensated by scale-size
(LHS of \eqref{K41precise.eq}) at the three $\Re$ shown in Figure \ref{fig1}({\textbf{A}}), plotted against non-dimensional scale $\ell/\L$. Horizontal lines show respective $(4/5) \epsn$  
(RHS of \eqref{K41precise.eq}) at the same three $\Re$.
The ordinate is non-dimensionalized by ${\U}^3/\L$.
Error bars in symbols and the shaded regions represent temporal standard deviations in $\srlong$ and $\epsn$ respectively, in statistically steady state.
With increasing $\Re$, $\srlong$ develops a linear scaling or the ratio $\srlong/\ell$ develops a plateau which matches up with $(4/5)\epsn$ over a widening scale-range extending towards smaller scales -- this is considered as the putative inertial range and is demarcated by filled symbols;
consistent with decreasing $\du$ with increasing $\re$ (indicated by dotted arrow), the plateau height decreases in the direction marked by the solid arrow such that 4/5th law \eqref{K41precise.eq} holds.  
}
\label{fig3}
\end{figure}

\vspace{0.2cm}
These results raise many questions, in particular the consistency of our results so far with the celebrated 4/5th law, which was 
derived by Kolmogorov \cite{kolmogorov1941dissipation} assuming \eqref{zerothlaw} and which has been confirmed
in high-$\Re$ experiments \cite{KRS98} and simulations \cite{iyer2020scaling}. The 4/5th law states that 
\be
\label{K41precise.eq}
 \lim_{\ell \to 0} \lim_{\Re \to 0} \frac{-\srlong}{\ell} = \frac{4}{5}  \lim_{\Re \to 0}\epsn
 \,
\ee
where $\srlong =\la (\drulong)^3 \ra_{\Omega,\ux,t}$ is the third-order longitudinal velocity structure function defined for the longitudinal increment 
$\drulong \coloneqq \delta_\br\bu(\bx,t) \cdot \ur/\ell$ at scale-size $\ell$; $\la \cdot \ra_{\Omega,\ux,t}$ denotes the average over solid angle, space as well as time. It has been shown, however, that the 4/5th-law remains valid 
even under the assumption that
\be\label{weakdiss}
 D[\bu]  \sim \Re^{-\alpha}, \quad \alpha<1\, \quad \mbox{ as } \Re\to\infty
\ee
which for $0<\alpha<1$ is labeled as {\it weak dissipative anomaly}, in contrast to the strong dissipative anomaly for
$\alpha=0$ \cite{bedrossian2019sufficient}. 
It is easy to check that the Taylor microscale 
defined by $\lambda^2 = 15\nu \U^2/\epsn$
satisfies $\lim_{\Re\to\infty} \lambda/\L=0$ precisely when {$\alpha<1$} in \eqref{weakdiss}, for which it is proved 
in \cite{bedrossian2019sufficient} that the maximum difference between $-\sr/\ell$ and $\frac{4}{5}\varepsilon$ 
becomes vanishingly small over an increasing range of length-scales $\lambda\ll \ell\ll \L.$ This result has the notable  
implication that validity of the 4/5th law cannot be taken as evidence for the zeroth law. The proof in \cite{bedrossian2019sufficient} 
considers the statistical steady-state of turbulence in a periodic domain driven by a body force which is a Gaussian random field,
delta-correlated in time and takes advantage of special features of that stochastic forcing. Recent work has established similar results in the deterministic setting \cite{novack2023scaling}.  Thus,  flows having smoothness above that required for anomalous dissipation do not necessarily contradict the 4/5th law, which -- we remark -- itself possesses certain regularizing properties \cite{drivas2022self}.
Finally, despite the fact that $S_3(\ell)\sim \ell^{\zith}$ with $\zith > 1$ in the inertial range, the inequality $|S_{3,\parallel}(\ell)| \leq S_3(\ell)$
does not contradict the 4/5th law 
$
-\srlong \sim \frac{4}{5} \varepsilon \ell.$ 
{Indeed, assuming that \eqref{weakdiss} holds with the exponent $\alpha$ from \eqref{expcond}} 
\begin{equation}\label{45bnd}
\frac{|S_{3,\parallel}(\ell)|}{\varepsilon \ell} \leq C_T \left( \frac{\U^3}{\varepsilon \L} \right) \left(\frac{\ell}{\L}\right)^{\zeta_3-1} \! \! \sim \!  \left({\Re}^{\frac{3}{3+ \zeta_3}}\frac{\ell}{\L}\right)^{\zeta_3-1}.
\end{equation}
 The upper bound only becomes order unity for 
  $\ell/\L \sim \Re^{-3/(3+\zeta_3)} $, which is essentially the Kolmogorov scale $\Re^{-3/4}$!  Thus, the divergence of  the prefactor 
 ${\U^3}/{\epsn \L}$ on the right-hand side of \eqref{45bnd} accommodates the decay in ${\ell}/{\L}$ within the inertial range, allowing these two results to coexist.

\vspace{0.2cm}
Although {our forcing is quite different from that considered in the proof of \cite{bedrossian2019sufficient}}, 
we may check numerically the validity of the 
4/5th law. 
With increasing $\Re$ and decreasing $\ell$, Figure \ref{fig3} shows that $-\srlong/\ell$ evidently matches $4\epsn/5$ over an increasing 
scale range (extending towards decreasing scale-sizes), which we identify as the inertial range. Also evident is the slight decrease in $\epsn \elo /{\U }^3$ 
(indicated by the downward arrow) with increasing $\Re$. 
Our numerical results are consistent with the conclusions of \cite{bedrossian2019sufficient} and fit with analogous theorems in the deterministic setting of \cite{novack2023scaling}. 

\vspace{0.2cm}
In conclusion, we have examined the zeroth law for homogeneous and isotropic turbulence in a periodic box in light of the 
Reynolds-number dependence of the third-order absolute velocity scaling exponents. These exponents show a persistent trend for $\zithnu$ towards a value slightly above unity with increasing $\Re$. In accord with this observation together with theory \cite{drivas2019onsager}, our results suggest that incompressible HIT in a periodic box exhibits a \emph{weak dissipative anomaly}, at least for the specific numerical set-up that we consider, {similar to what is observed in smooth-wall pipes or Taylor-Couette flows}. 
We cannot exclude the occurrence of a qualitative shift in the scaling at some higher $\Re_*$ with $\zithnu \to 1$, consistent with the zeroth law, making the present range effectively transient \footnote{Analogous longitudinal exponents from experiments \cite{KRS98,Dhruva2000} at Taylor scale Reynolds number of $10,340$ are smaller than our extrapolated asymptotic $\zith$, and the exponent from \cite{Dhruva2000} is smaller than the $\zithnu$ measured in the present DNS at this largest Reynolds number at which such measurements have been made, although larger than unity (it is reported as $1.03\pm 0.03$).  See Figure 3 of the Supplemental Information. These measurements were made in the atmospheric boundary layer with large-scale variability, and cannot be regarded as fully satisfactory. Nevertheless, they suggest that the asymptotic $\zith$ may, in fact, be smaller than our estimate here, and leaves open the possibility that $\zith \le 1$.
}. We can, however, assert that our results in the range of Reynolds numbers explored have converged and that the small-scales satisfy the exact relations (4/5-ths and 4/3-rds laws) with increasing fidelity as $\Re$ increases, and that basic measures of local isotropy expected in HIT flows have been met. We also observe that the present result does not imperil the Kolmogorov 4/5-ths law.
Furthermore, a strong dissipative anomaly remains consistent with empirical evidence 
for turbulent flows with solid boundaries, particularly external wakes \cite{sreenivasan1984scaling,pearson2002measurements} 
and internal flows with rough walls \cite{nikuradse1933laws,cadot1997energy}.
Obviously, given the importance of our result, further investigation is called for -- not only for numerical simulations 
but also for a wide variety of laboratory flows. 

\section{Methods}
\textbf{Direct Numerical Simulations:} The 3D incompressible Navier-Stokes equations are numerically solved using a pseudo-spectral solver in a periodic cube with edge-length $2\lbox$ with $N$ grid points per side; the uniform grid-spacing $\Delta x = 2\lbox/N$. Reynolds number $\Re$ is increased by decreasing the kinematic viscosity $\nu$ over successive simulations and increasing $N$ proportionally. 
The grid-resolution $\Delta x/\eta $, $\eta \equiv (\nu^3/\epsn)^{1/4}$ being the average Kolmogorov length scale, is $1.5$ at the highest $\Re$ simulated.
Time advancing is done using a second-order Runge-Kutta scheme. Turbulence is maintained in a statistically stationary state by freezing the energy spectrum for small wavenumbers at long-time averages (see \cite{SIpaper} for some details) from prior simulations \cite{DY10}. 
The velocity field $\uu$ is self-similar across all $\Re$ recorded \cite{KI23}. 
The DNS data are saved periodically over eight eddy-turnover times $(\sim \L/\U)$ for calculating statistical averages. See Refs.~\cite{KIKRS18,iyer2020scaling} for details on the small-scale resolution and convergence studies for this DNS database.
\vspace{0.2cm}

\textbf{Third-order scaling:} The structure functions $\srv$ are calculated along direction ${\ur}/\ell$ by measuring the two-point velocity increments $\druv$ for all space-points $\ux$ at a given time $t$. The isotropic projection 
$\sr$ is then obtained by angle averaging as detailed in \cite{rkf17,iyer2020scaling}. Although the large scales $O(\L)$ inherit the specifics of our numerical set-up, the small-scales $\ell/\L \ll 1 $ are increasingly isotropic with increasing $\Re.$ 
Expectations of isotropy in HIT (at least up for statistical moments up to order four) are satisfied increasingly closely with increasing $\Re$ as shown in Ref.~\cite{iyer2020scaling} and discussed in the Supplementary Information \cite{SIpaper}. 
We show there that the third-order exponents presented in this work are not sensitive to the (finite) temporal and spatial resolution effects of the DNS, especially at higher $\Re$. Lastly, the simulation domain size $2\lbox$ and specifics of the large-scale forcing have been verified \cite{SIpaper} to have no effect on the measured exponents at higher-$\Re$ presented in Figure \ref{fig1}. 


\section{Acknowledgments}
KI's work was supported by a Research Seed award from the Office of the Vice President for Research at Michigan Tech.
The research of TDD was partially supported by the NSF DMS-2106233 grant, NSF CAREER award \# 2235395, Stony Brook University Trustee’s award and Alfred P. Sloan Fellowship. 
This work used Stampede3 at Texas Advanced Computing Center (TACC) at The University of Texas at Austin through allocation PHY200084 from the Advanced Cyberinfrastructure Coordination Ecosystem: Services \& Support (ACCESS) program, which is supported by U.S. National Science Foundation grants \#2138259, \#2138286, \#2138307, \#2137603, and \#2138296. KRS's research was supported by New York University. We are grateful to P.K. Yeung for long-standing collaboration on DNS data.

\bibliography{bibliography}
\end{document}


\title{Supplementary information on ``Whither the Zeroth Law of Turbulence?"}
\author{Kartik P. Iyer}
\affiliation{Department of Physics and Department of Mechanical and Aerospace Engineering, Michigan Technological University, Houghton, MI 49931, USA}
\email{kiyer@mtu.edu}
\author{Theodore D. Drivas}
\affiliation{Mathematics Department, Stony Brook University, Stony Brook, NY 11794, USA}\email{tdrivas@math.stonybrook.edu}
\author{Gregory L. Eyink}
\affiliation{Department of Applied Mathematics and Statistics, and Department of Physics, The Johns Hopkins University, Baltimore, MD 21218, USA}
\author{Katepalli R. Sreenivasan}
\affiliation{Department of Physics, and the Courant Institute of Mathematical Sciences, New York University, New York, NY 11201, USA}

\maketitle
\noindent
This supplementary information provides additional details reinforcing the main text, though it is self-contained for the most part.
\begin{figure}
\centering
\includegraphics[width=0.5\columnwidth]{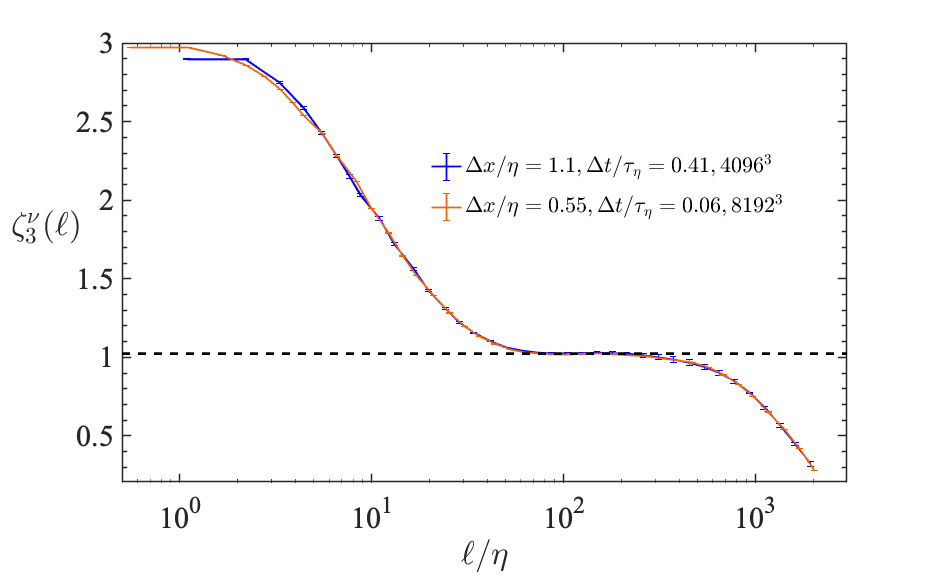}
\protect\caption{
Logarithmic local slope of the angle-averaged third-order absolute velocity structure function $\sr$ \eqref{lstot.eq} {\it{vs.}}~scale $\ell/\eta$, $\eta$ being the Kolmogorov scale.
The two curves correspond to two simulations on periodic boxes with edge-lengths $2\lbox = 2\pi$ with same viscosity ($\re \approx 3 \times 10^4$, $\eta/\lbox = 4 \times 10^{-4}$), 
but with different spatial and temporal resolutions; 
$\Delta x = 2\lbox/N$ is the uniform grid-spacing, $\Delta t$ is the time-step and
$\tau_\eta$ is the Kolmogorov time scale.
The limit $\zithnu(\ell) \to 3$ from first-order Taylor expansion is more evident in the finer DNS.
In the inertial range, however, $\zithnu(\ell)$ for both simulations collapse to the same 
$\ell$-independent plateau which is the measured scaling exponent $\zithnu$. Power-law exponent obtained from a least-squares fitting to $\sr$ in the inertial range (dashed line) agrees well with the local slope exponent.
}
\label{res.fig}
\end{figure}

\section*{Third-order absolute scaling} 
Scaling exponents of the third-order absolute structure functions are calculated by 
power-law fits to the inertial range (determined by the Kolmogorov 4/5th law as shown in the main text) by the method-of least-squares. 
Two additional methods to determine the exponents are now provided. {Angle-averaging of the 
structure functions has been performed as stated in the text, in order to improve scaling. Note 
that the mathematical estimate of \cite{drivas2019onsager} was derived using scaling exponents for 
structure functions without angle averaging. However, it is easy to modify their proof by requiring filter 
kernels $G({\bf \ell})$ to be spherically symmetric, depending only on $\ell=|{\boldsymbol {\ell}}|,$ so that the 
established bounds involve the exponents of the angle-averaged structure functions.}
\begin{figure}[h!]
\includegraphics[width=0.47\columnwidth]{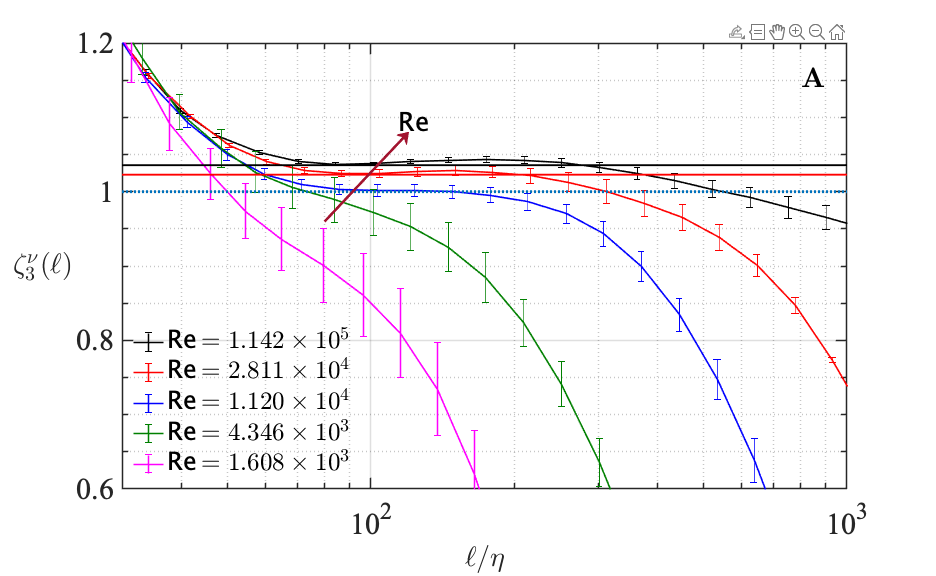}
\includegraphics[width=0.47\columnwidth]{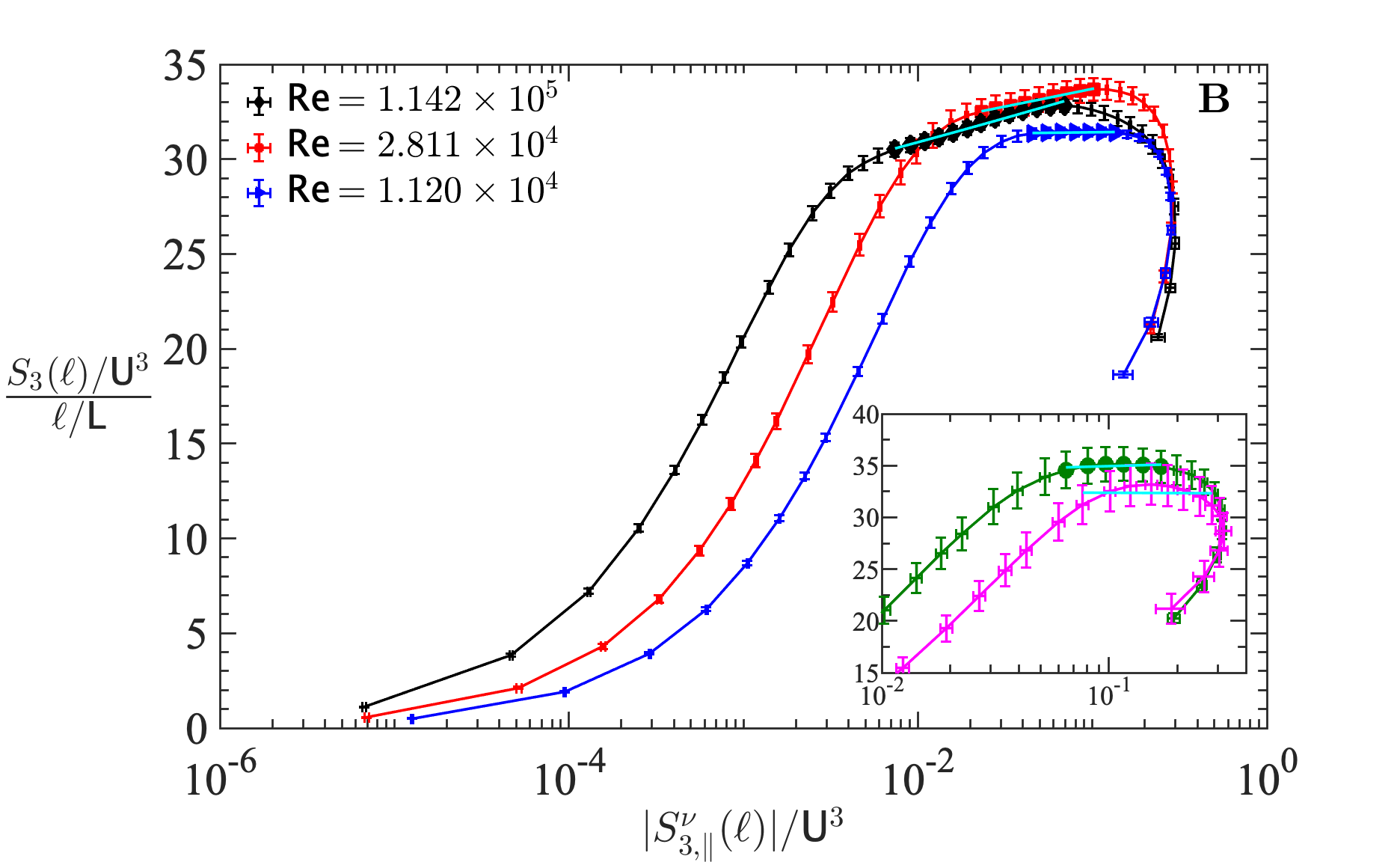} \\
\includegraphics[width=0.45\columnwidth]{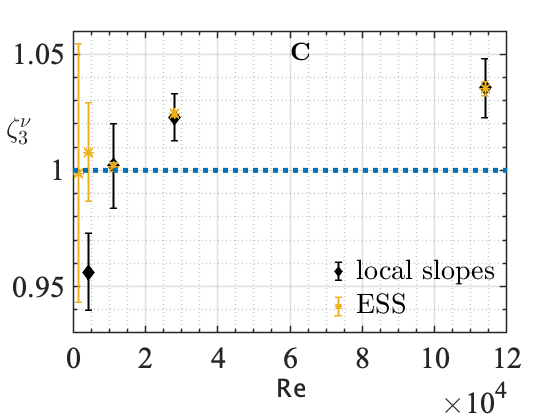}
\protect\caption{
Reynolds number trends of the inertial range scaling exponents $\zithnu$ obtained from (A) local slopes \eqref{lstot.eq} and (B) method of Extended Self-Similarity (ESS). 
At low $\Re$, local slopes show no discernible scaling as seen in (A), whereas method of ESS yields some scaling as seen in inset of (B).
At higher $\Re$ for which exponents can be determined from both methods, they agree as seen in panel (C).
Solid horizontal lines for the three highest $\Re$ in (A) correspond to $\zithnu$ from inertial range power-law fits from Figure $1$ of main text to demonstrate agreement between power-law fits and local slopes. Error bars correspond to temporal fluctuations in the statistically steady state.}
\label{totexp.fig}
\end{figure}


{\vspace{0.2cm}
Logarithmic local slopes of the third-order absolute structure function $\sr$ defined as
\beq
\label{lstot.eq}
    \zithnu (\ell) = \frac{d [\log \sr]}{d[\log \ell]} \;, \\
\eeq
are shown in Figure \ref{res.fig} at a fixed Reynolds number $\re$. In the intermediate scale range $1 \ll \ell/\eta \ll \L/\eta$,
$\zithnu(\ell)$ assumes a scale-independent plateau which is the inertial range scaling exponent $\zithnu$. 
The range
where the third-order scaling is detected is also the same range where Kolmogorov 4/5th law holds (see Figure $3$ of main text). The scaling exponent obtained by a power-law fit to the inertial range (from Figure $1$ in main text) is also verified to be consistent with the local slope exponent in Figure \ref{res.fig}. 
Furthermore, by using a shorter DNS with finer space and time resolution we have verified in Figure \ref{res.fig} that $\zithnu$ reported here are not affected by resolution artifacts. 

{\vspace{0.2cm}
Figure \ref{totexp.fig} contrasts the estimation of $\zithnu$ using local slopes \eqref{lstot.eq} with the method of extended self-similarity -- abbreviated as ESS \cite{benzi93}, at different $\re$. 
The ESS scaling range marked by symbols in panel \textbf{B} is also the range where $\srlong \sim \ell$ -- which is taken as the putative inertial range. Power-law least-squares fitting in this range has been used to quantify the deviation from linear scaling.
With increasing $\re$ the ESS scaling appears to increase beyond unity, in a manner that is consistent with the local slopes shown in panel \textbf{A}. 
At lower $\re$ the scaling is restricted to a narrow range and appears to be approximately linear, with the narrow fitting range at low-$\re$ rendering the results extremely sensitive to the exact choice of the ``inertial range". As a result, the low-$\re$ exponents are typified by large error bars as shown in Figure \ref{totexp.fig}. However, with increasing $\re$ the inertial range extent increases; as a result, the empirically determined exponents become sensibly independent of the particulars of the extraction method, or to slight changes in the choice of the inertial range. Panel \textbf{C} in Figure \ref{totexp.fig} illustrates the robustness of the high-$\re$ third-order scaling exponents $\zithnu$ to the method used to quantify them. 

\clearpage 

\section*{Asymptotic third-order exponent $\zith$}
The asymptotic third-order exponent is obtained from the DNS exponents $\zithnu$ by curve fitting using the method of least-squares. In Figure \ref{fits.fig} we have shown three different fits including both possibilities $\zith \le 1$ ($\dstar > 0$) and 
$\zith > 1$ ($\dstar = 0$). While all the three fitting functions fit the DNS data well, the fit standard error (RMSE) for both the $\dstar = 0 $ cases shown are $30 \%$ smaller relative to the $\dstar > 0$  fit. Table \ref{fits.tab} also shows that the spread in the asymptotic fit parameter $\zith$ is smaller in the logarithm modulated power-law than the pure power-law fit and is hence considered as the best fit to the data.

\begin{figure}[h!]
\begin{minipage}[b]{1.0\linewidth}
\includegraphics[width=0.6\linewidth]{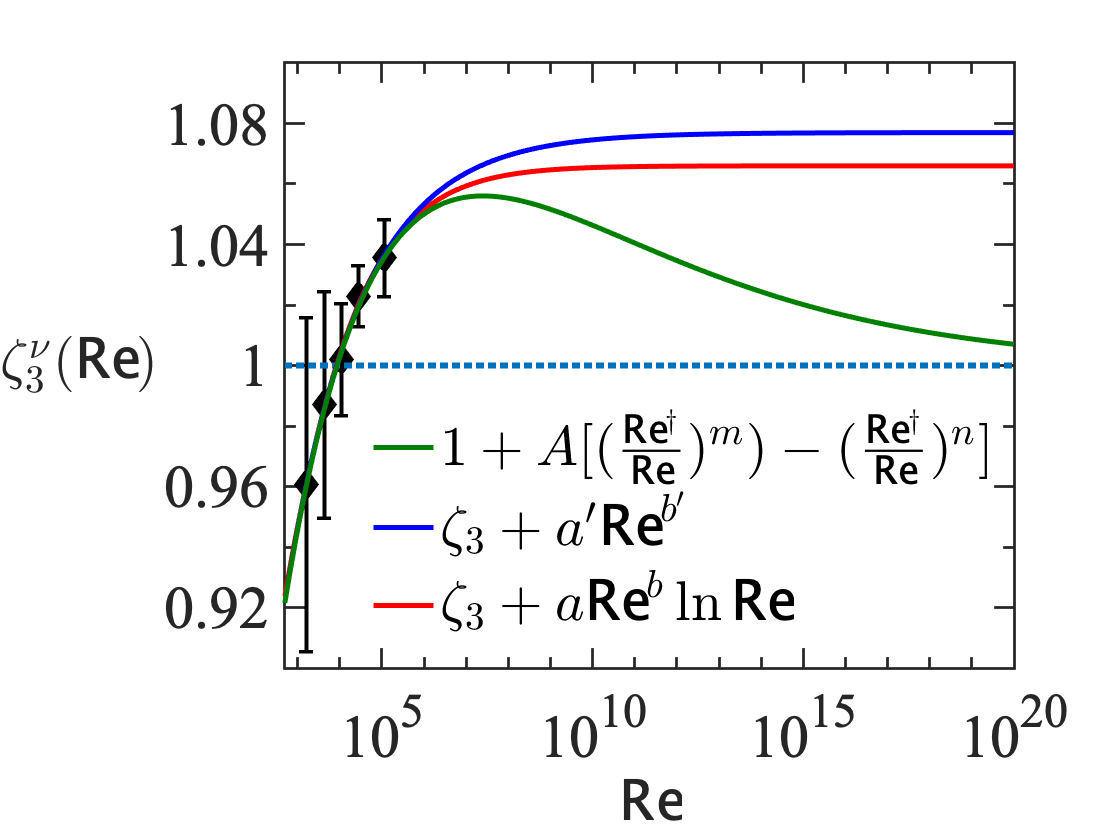}
\captionof{figure}{Estimation of third-order asymptotic exponent using method of least-squares. Symbols correspond to $\zithnu$ from DNS, solid lines are three different fits to the DNS exponents. Fit details provided in Table \ref{fits.tab}.}
\label{fits.fig}
 \end{minipage} 
 \vfill
 \begin{minipage}[b]{1.0\linewidth}
\begin{tabular}{|c|c|c|c|c|} \hline
   fit function &  fit parameters & $\zith$ &RMSE & asymptotic dissipation \\
     \hline
 $1+A \big [\big (\frac{\mathsf{Re^\dagger}}{\mathsf{Re}} \big )^m-\big (\frac{\mathsf{Re^\dagger}}{\mathsf{Re}} \big )^n \big ]$ & $A = 0.631$, $\mathsf{Re^\dagger} = 8634$, $m = 0.111$, $n  = 0.141$ & $1$ & $0.0047$& $\dstar >0$ \\
$\zeta_3 + a^{\prime} {\mathsf{Re}}^{b^{\prime}}$& $\zith = 1.077$,$a^\prime=-0.724$, $b^\prime = -0.248$ & $1.077 \pm 0.1110 $ &$0.0033$ & $\dstar = 0$ \\
$\zeta_3 + a {\mathsf{Re}}^b \ln {\mathsf{Re}} $ & 
$\zith  = 1.066 $, $a = -0.284$,$b =-0.406$ & $1.066 \pm 0.069$ & $0.0032$ & $\dstar = 0$ \\
\hline
\end{tabular}
\captionof{table}{Details of the fits shown in Figure \ref{fits.fig}; fit parameter $\mathsf{Re}^\dagger$ roughly corresponds to $\rel \approx 400$ where $\zithnu \approx 1$.
Root Mean Squared Error (RMSE) of the fits are provided; $\dstar$ follows from $\zith$ by theory \cite{drivas2019onsager}.}
\label{fits.tab}
 \end{minipage}
\end{figure}

\clearpage 

\section*{Third-order absolute longitudinal scaling}

Since the 
third-order absolute longitudinal exponents
$\la |\drulong|^3 \ra \sim \U^3 (\ell/\L)^{\zithnulong}$ have been measured in both simulations and experiments, we report the same from our DNS in Figure \ref{long.fig}. The qualitative trend of $\zithnulong$ with increasing $\Re$ appears the same as that of $\zithnu$ as shown in panel \textbf{A} which shows the local slopes. The longitudinal exponents from our simulations are compiled along with past measurements in panel \textbf{B} of Figure \ref{long.fig} over four decades of Taylor-scale Reynolds number $\rel$. Since the large length-scale $\L$ is different for different studies, $\rel$ is a more convenient choice for such a comparison. The longitudinal exponents reported in  \textbf{B} 
correspond to different studies of HIT each with its own peculiarities (see Tables \ref{tab1} and \ref{tab2}). In spite of the appreciable error bars, the general trend from different measurements appears to suggest that $\zithlong = \lim_{\nu \to 0} \zithnulong > 1$.
We especially take note of the
two overlapping measurements of $\zithnulong$
from atmospheric boundary layer experiments at $\rel = 10,340$ reported in Refs.~\cite{KRS98,Dhruva2000}. 
These correspond to the highest $\rel$ reported up-to-date. 
Both these measurements show considerable scatter, but overwhelmingly support $\zithlong >1$ with that of Ref.~\cite{Dhruva2000} including the possibility that $\zithlong = 1$.  It is hard to fit those measurements neatly into the category of HIT data considered here, but they do not rule out the possibility that $\zithnulong$ will asymptotically approach unity.

\begin{figure}[h!]
\includegraphics[width=0.4973\columnwidth]{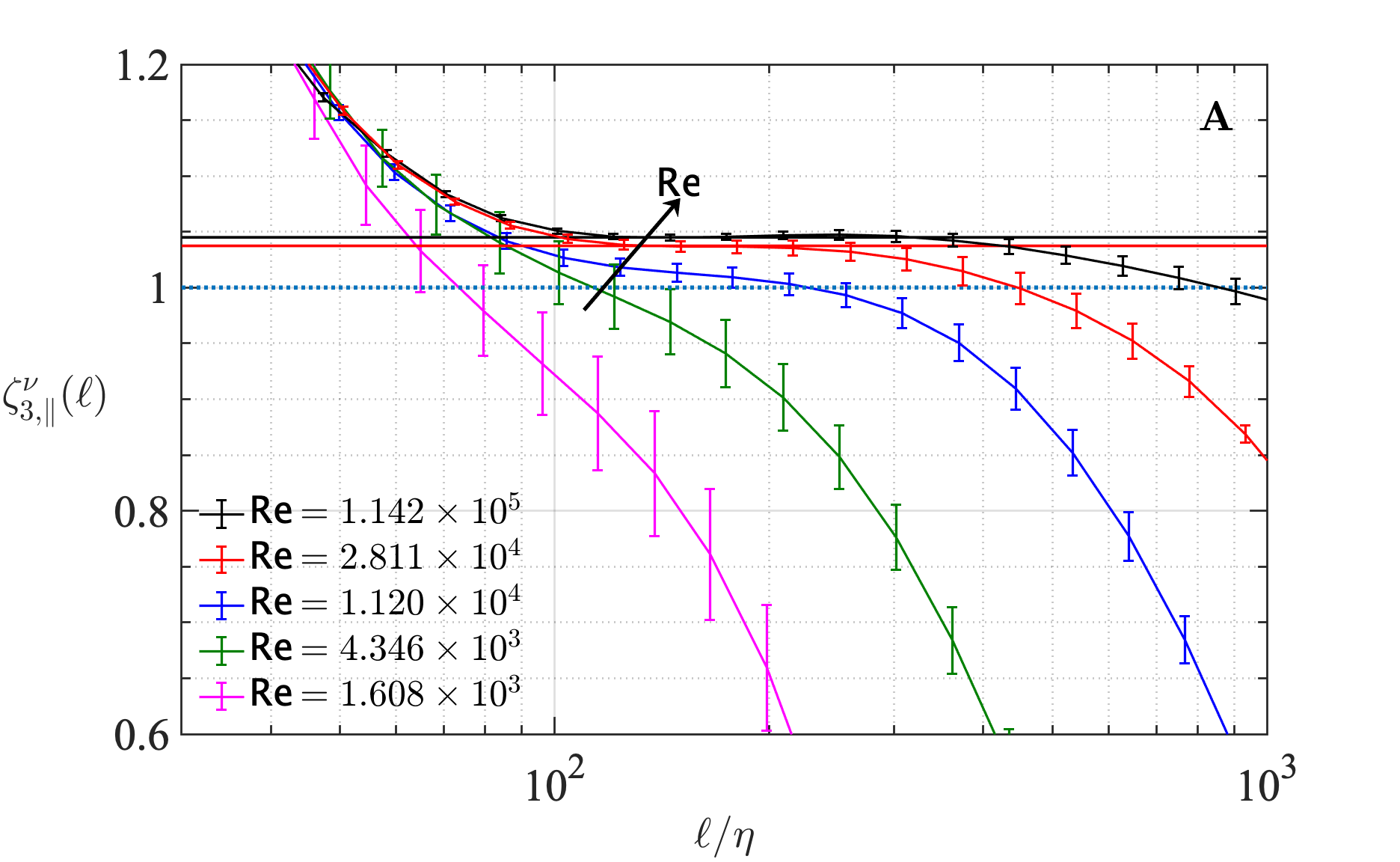}
\includegraphics[width=0.4973\columnwidth]{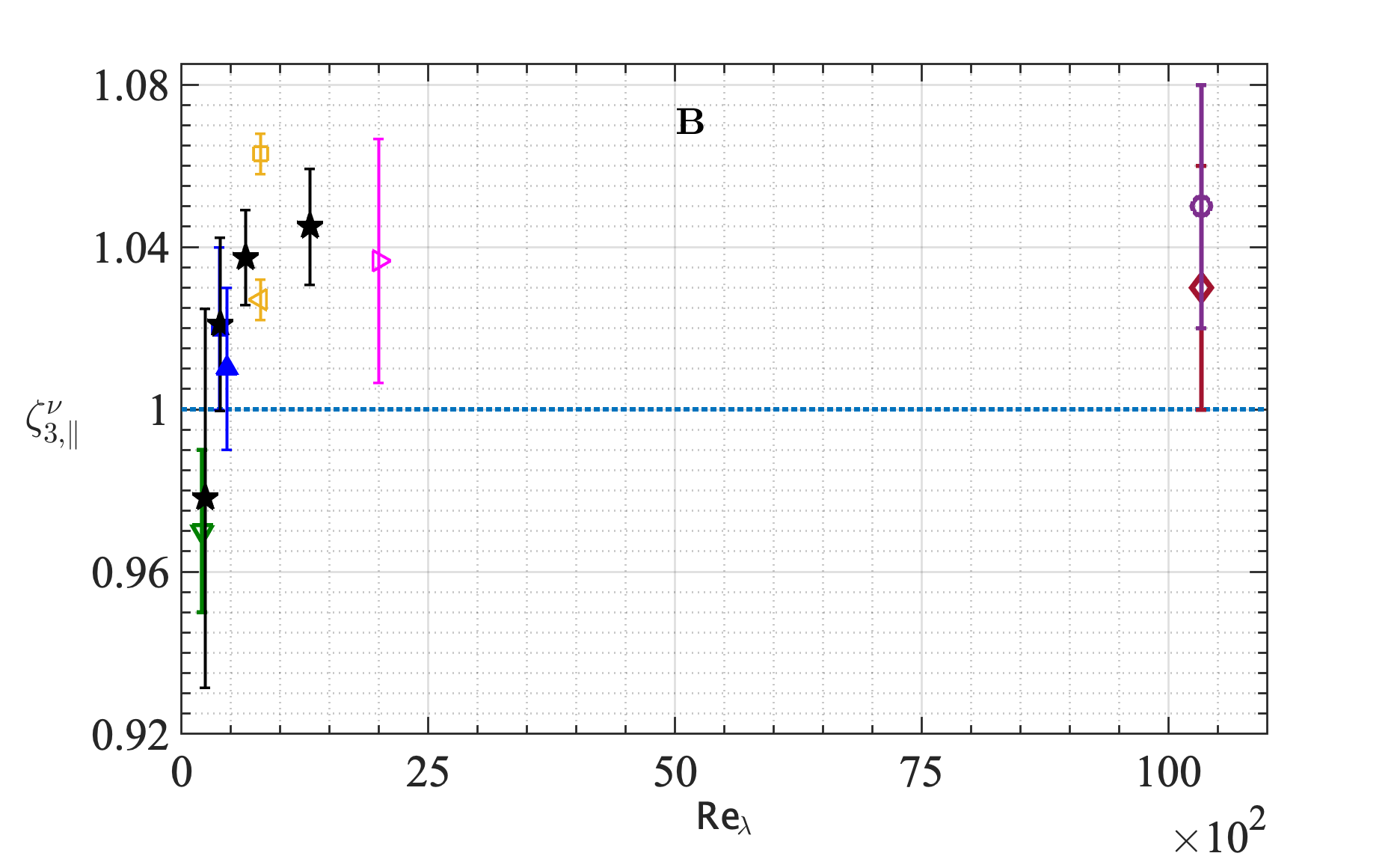}
\protect\caption{Reynolds number evolution of the absolute longitudinal exponents. (A) Local slopes 
$\zithnulong (\ell) = {d [\log \la |\drulong|^3 \ra]}/{d[\log \ell]} 
$ plotted against $\ell/\eta$ at different $\re$. With increasing $\re$ the local slopes assume $\ell$-independent plateaus which are the assumed inertial range longitudinal absolute scaling exponents $\zithnulong$. Solid horizontal lines are $\zithnulong$ obtained from power-law fits in the region where Kolmogorov's 4/5-ths law holds and are consistent with the local slope exponents as shown.
(B) Compilation of  $\zithnulong$ $({{\bigstar}})$ from panel {\textbf{A}} as a function of Taylor-scale Reynolds number $\rel$. Also plotted are $\zithnulong$ from other simulations (filled symbols) and experiments (unfilled symbols), see Tables \ref{tab1} and \ref{tab2} for details. 
}
\label{long.fig}
\end{figure}


\begin{table}[h!] 
\centering
    \begin{tabular}{|c|c|c|c|c|c|c|c|} \hline
    &  space average & method & force & forcing range &$T_E$  & Reference\\
     \hline
 $({{\bigstar}})$ & angle average & local slope, power-law & frozen-energy& $0 < k \le 2.1$ &$8$  & 
 present \\
$(\colb{\blacksquare})$& angle average & local slope & Gaussian random white-in-time& $1 \le k \le \sqrt{6}$  &$4.21$  &
 \cite{GFN02} \\
$(\colb{\blacktriangle})$ & angle average & local slope & Gaussian random white-in-time& $1 \le k \le \sqrt{6}$  &$2.14$ & 
 \cite{GFN02} \\
     \hline
    \end{tabular}
   \caption{Summary of third-order absolute exponents $\zithnulong$ from DNS of HIT.
   All simulations are three-dimensional, simulated in periodic cubes with edge-length $2\lbox=2\pi$; ``space average" is the technique used to calculate the structure functions $\langle |\drulongonly|^3\rangle$;  exponents were extracted from $\langle |\drulongonly|^3\rangle$  using ``method"; forcing is applied in Fourier space at wave-number magnitudes $k$, $T_E = \lint/\up$ \eqref{lscales.eq} is the large-eddy turnover time in statistically steady-state used for averaging purposes. 
   }
    \label{tab1}
\end{table}
\begin{table}
\centering
    \begin{tabular}{|c|c|c|c|c|} \hline
   &  data   & samples & method  &Reference\\
     \hline
     $(\colg{\triangledown})$  
     & atmospheric boundary layer   & $10^7$
     & power-law & \cite{KRS94} \\
$(\colm{\triangleright})$ & low temp Helium  & $3 \times 10^6$
& power-law & \cite{Belin96} \\
$(\colbr{\ast})$ & atmospheric boundary layer   & $ 10^7$
& ESS & \cite{Dhruva2000} \\
$(\colv{\circ})$ & atmospheric boundary layer  & $10^7$
& ESS & \cite{KRS98} \\
 $(\colo{\triangleleft})$ & jet centerline  & $10^9$
& ESS &\cite{Herweijer} \\
$(\colo{\Box})$ & jet centerline & $10^9$
& power-law & \cite{Herweijer} \\
     \hline
    \end{tabular}
   \protect\caption{Summary of third-order absolute exponents $\zithnulong$ from experiments. The experimental set-up is given by  ``data"; 
   ``samples" refers to the temporal data records that were converted into one-dimensional cuts of the structure function using Taylor"s hypothesis for space-time surrogacy;
   ``method'' refers to the procedure used to obtain $\zithnulong$ from the structure functions.
  }
    \label{tab2}
\end{table}

\clearpage 

\section*{Local Isotropy} 
An assessment of isotropy at three vastly different scales from the periodic DNS with box length $2\lbox$ is provided in Figure \ref{isoscale.fig}. 
At the largest scale permissible, the one-dimensional velocity correlations (both longitudinal and transverse) do not decay with increasing $\re$ because of residual anisotropy from the large-scale forcing and cubic grid geometry, at the largest scales \cite{ISY17}. In contrast, the corresponding three-dimensional velocity correlations which involve both longitudinal and transverse contributions, show the expected HIT behavior -- presumably because the anisotropy between the longitudinal and transverse components compensate one-another in a manner that is consistent with HIT.
In order to assess whether this large-scale anisotropy persists down-scale we plot the ratio $g(\ell)$ of the right-hand-side and the left-hand-side of the isotropic incompressible relation 
\beq
\label{secondiso.eq}
\srtranstwo = 2\srlongtwo + \ell \frac{d}{d\ell} \srlongtwo \;,
\eeq
at an intermediate scale which resides in the inertial range
in Figure \ref{isoscale.fig}(B). Here, 
$\srlongtwo = \langle (\drulong)^2  \rangle$ is the second-order longitudinal velocity structure function 
and 
$\srtranstwo = \langle |\druperp|^2  \rangle$ 
is the second-order transverse velocity structure function; the longitudinal and transverse velocity increments are related to the total increment,
$\druv = \druperp + \drulong \widehat{\boldsymbol{\ell}}$. 
The ratio $g(\ell)$ for the two arbitrary directions $\bf{\ell}$ shown in Figure \ref{isoscale.fig} (B) do not tend to the isotropic value of unity with increasing $\rel$ due to the remnant anisotropy passed down-scale from the large-scale set-up of the DNS. Only the angle averaged isotropic value shows excellent correspondence with unity \cite{iyer2020scaling}. 
Finally, in panel (C) of Figure \ref{isoscale.fig} we verify the isotropic relation between the longitudinal and transverse velocity gradients, which correspond to the smallest scales computed in the DNS.
The conclusions from examining isotropic relations at the three disparate scales in the forced $2\pi$-DNS is that, although the large scales show behavior inconsistent with HIT, isotropy is increasingly recovered downscale with increasing $\Re$. In particular, the angle-averaged intermediate range statistics shown here are not sensitive to the large-scale set-up. Further checks on the temporal and spatial accuracy of the structure functions from the present DNS data are provided in \cite{iyer2020scaling}. 

\begin{figure}[h!]
\includegraphics[width=0.6\columnwidth]{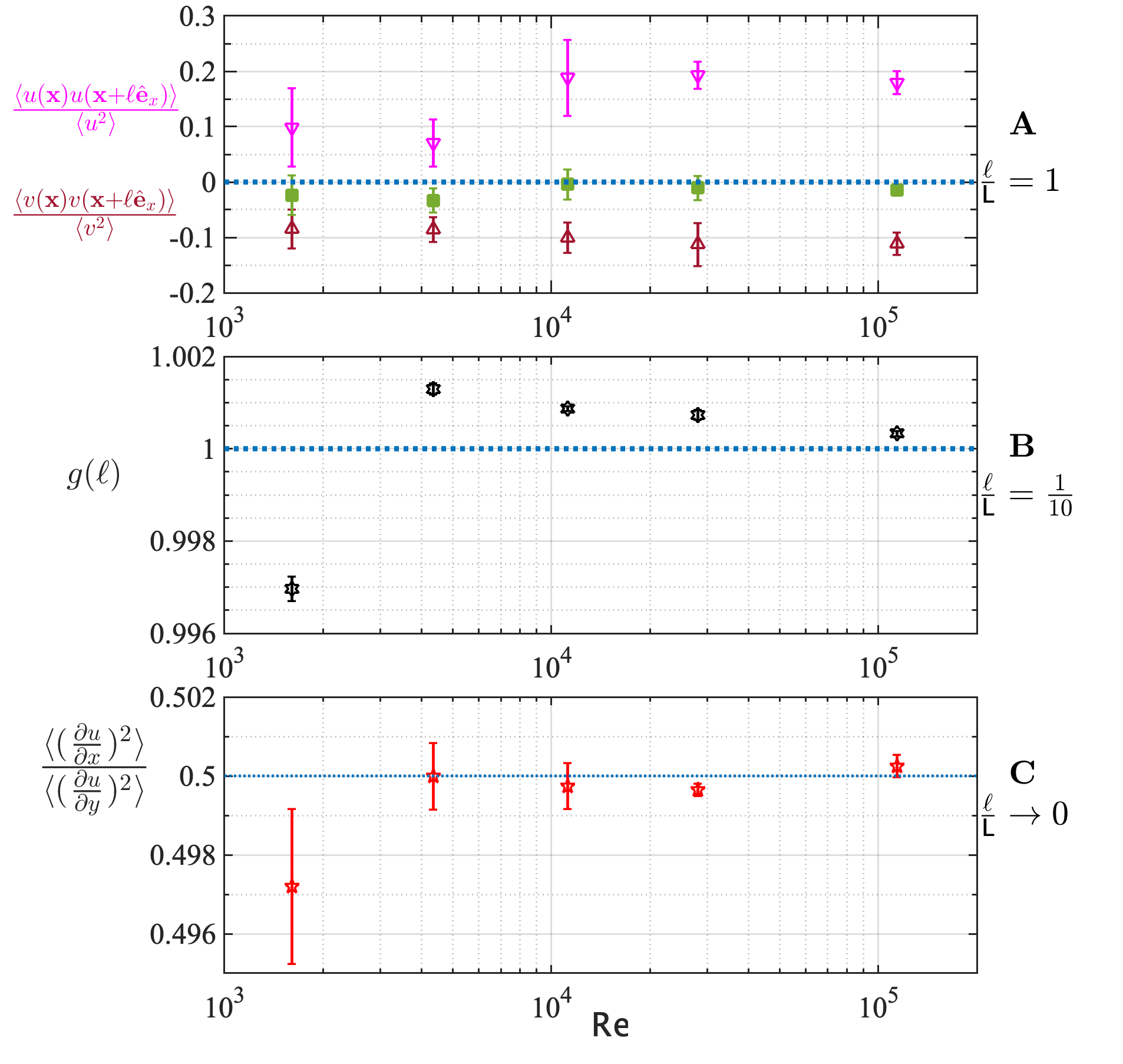} 
\protect\caption{Isotropy measures in periodic DNS with side-length $2\lbox = 2\pi$ plotted as a function of $\re$ at three disparate scales decreasing from panel (A) to panel (C). (A) One-dimensional longitudinal (down-triangles) and transverse (up-triangles) velocity correlations plotted at the largest scale possible which is the half-box length.  Also plotted are the corresponding three dimensional correlations 
$\la \uu(\ux)\cdot \uu(\ux+\ur) \ra/\la |\uu|^2\ra$ (filled squares). Dotted line at zero is the isotropic expectation.
(B) The isotropic ratio $g(\ell)$ defined in 
\eqref{secondiso.eq} at a smaller scale in the inertial range.
Open symbols correspond to separation lengths $\ell$ along arbitrary directions $\hat{\ur} = (0,1,0)$ (left-triangle) and $\hat\ur = (0,0,1)$ (right-triangle) whereas filled-stars corresponds to the isotropic angle average. 
Dotted line at unity is the isotropic value.
(C) Ratio of the longitudinal to the transverse velocity gradients which corresponds to the smallest scales calculated in the DNS. Dotted line at $0.5$ is the exact isotropic value.
}
\label{isoscale.fig}
\end{figure}

\begin{figure}[h!]
\centering
\includegraphics[width=0.6\columnwidth]{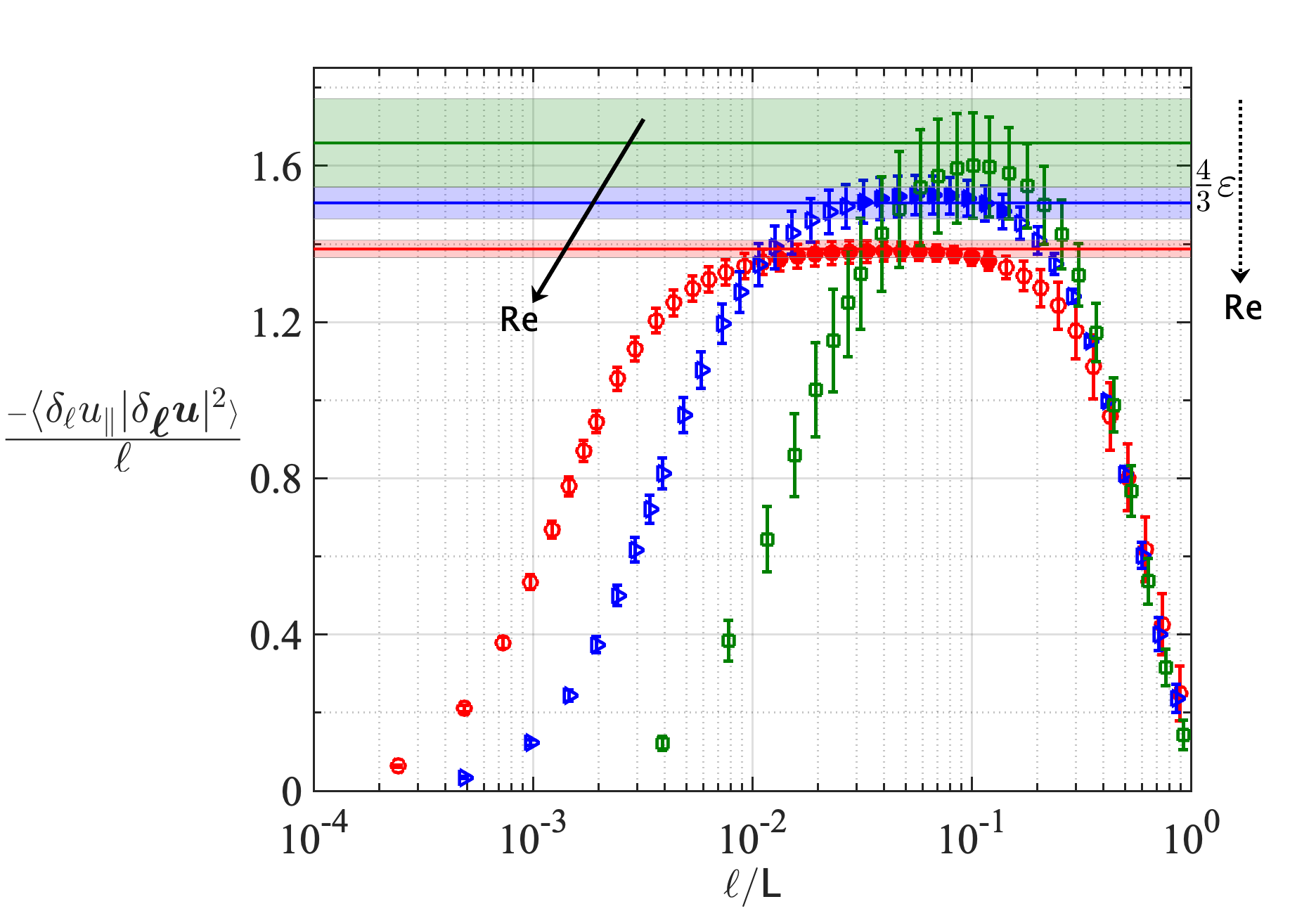}
\protect\caption{Verification of the angle averaged version of the 4/3rd law:
symbols correspond to the mixed structure function (LHS of \eqref{K43precise.eq}) at the three different $\Re$ shown in Figure $3$ of main text, plotted against non-dimensional scale $\ell/\L$. Horizontal lines show respective $(4/3) \epsn$  
(RHS of \eqref{K43precise.eq}) at the same three $\Re$.
The ordinate is non-dimensionalized by ${\U}^3/\L$.
Error bars in symbols and the shaded regions represent temporal standard deviations in $\langle \drulongonly |\druvonly|^2 \rangle$ and $\epsn$ respectively, in statistically steady state.
With increasing $\Re$, $\langle \drulongonly |\druvonly|^2 \rangle$ develops a linear scaling or the LHS of \eqref{K43precise.eq} develops a plateau which matches up with $(4/3)\epsn$ over the same widening scale region over which the 4/5th law holds which is the inertial range; consistent with decreasing $\du$ with increasing $\re$ (indicated by dotted arrow), 
the plateau height decreases with $\re$ (in the direction shown by solid  arrow), consistent with 
{Figure 3 in the main text where, likewise, $\du$ decreases} such that 4/5-ths law \eqref{K43precise.eq} holds. 
}
\label{k43.fig}
\end{figure}

{\vspace{0.2cm}}
As specific checks on our DNS we show in Figure \ref{k43.fig} the numerical verification of the 4/3-rds law which states that
\be
\label{K43precise.eq}
 \lim_{\ell \to 0} \lim_{\Re \to 0} \frac{-\langle \drulongonly |\druvonly|^2 \rangle_{\Omega,\ux,t}}{\ell} = \frac{4}{3}  \lim_{\Re \to 0}\epsn
\;.
\ee 
The angle averaged mixed structure function
$\langle \drulongonly |\druvonly|^2 \rangle$ plotted on the ordinate in Figure \ref{k43.fig} involves both the longitudinal and the total (or transverse) structure functions -- thereby providing a numerical verification of the total structure function data presented in this work. Note from Figure \ref{k43.fig} and from Figure \ref{isoscale.fig} (panel {\textbf{B}}) that the isotropic relation \eqref{secondiso.eq} is increasingly satisfied with $\re$ in the inertial range.
\begin{figure}[h!]
\centering
\includegraphics[width=0.5\columnwidth]{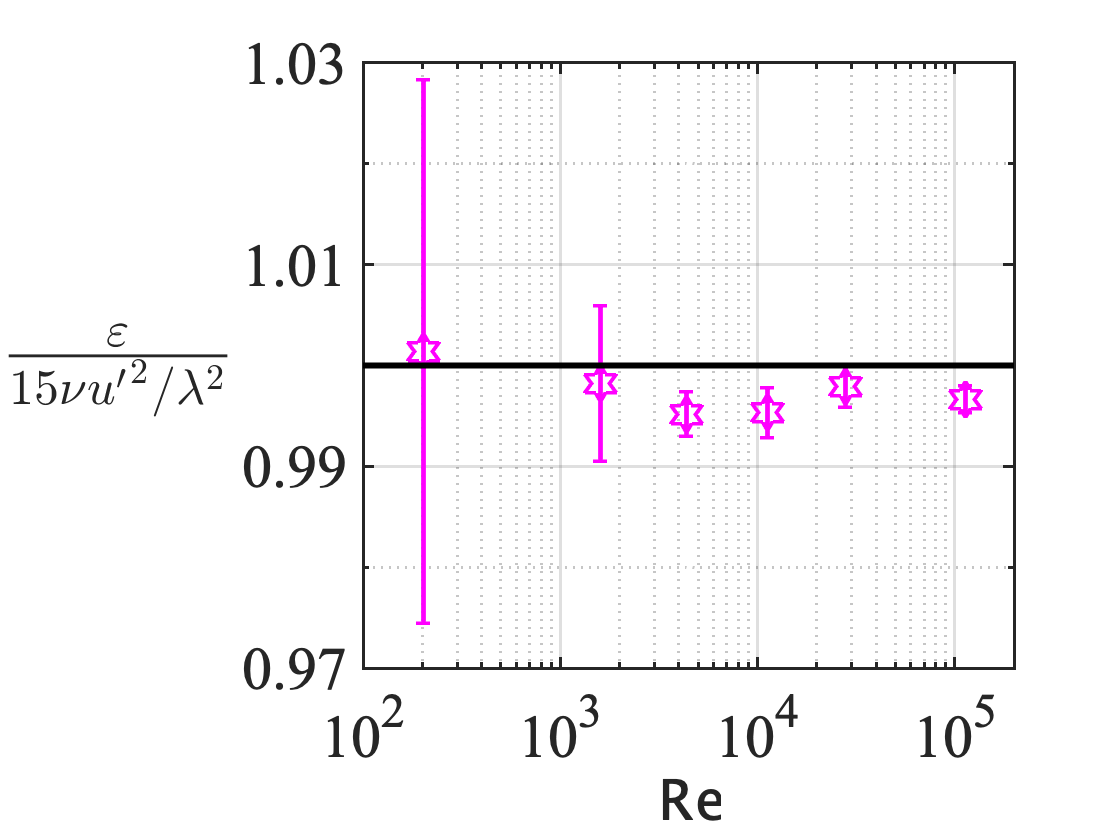}
\protect\caption{Ratio of the left-hand-side to the right-hand-side of \eqref{isochk.eq} as a function of $\rel$ in the
DNS database. Solid line at unity corresponds to perfect isotropy. With increasing $\re$ the error bars which are calculated as the standard deviation of the temporal fluctuations in the steady-state, substantially decrease, and are of the order of the symbol size at higher-$\re$.
}
\label{isochk.fig}
\end{figure}

{\vspace{0.2cm}}
Finally, the isotropic relation 
relation
\beq
\label{isochk.eq}
\epsn = {15 \nu {\up}^2}/{{\lambda}^2} \;,
\eeq
where $\lambda \coloneqq \sqrt{2} \up\la ({\partial u}/ {\partial y})^2\ra^{-1/2}$ is the Taylor microscale, is  examined in Figure \ref{isochk.fig}. At higher Reynolds numbers the ratio of the left-hand-side to right-hand-side is unity (which is the isotropic expectation) with better than $99\%$ accuracy.

\clearpage 

\section*{Box-periodicity and forcing} 
In order to assess the effects of the periodic boundary conditions and the large-scale forcing on the inertial range scaling presented in this work, we compare in each of the three panels of Figure \ref{boxlength.fig} two simulations at a fixed kinematic viscosity but with different box sizes, the second DNS with size double the first (and thus also with 
$Re$ doubled). The first pair of simulations (first column) is at $\rel=140$, second pair (middle column) is at
$\rel=240$ while the last pair (third column) is at $\rel=400$ in Figure \ref{boxlength.fig}.
Although the one-dimensional correlations differ at the largest scales for different box sizes as seen in the top panels, this difference is mitigated in the third-order structure functions which involves velocity differences (not shown).  It is evident from the bottom panels of Figure \ref{boxlength.fig} that the local slopes for the smaller and the larger box collapse better in the intermediate scale-range (around $\normr \approx 100$) with increasing Reynolds number -- presumably because of the increasing separation of the intermediate scales from the largest scales (where forcing and boundary conditions manifest) and the smallest scales (which are dominated by grid resolution and viscous effects).

\begin{figure}[h!]
\includegraphics[width=0.32\columnwidth]{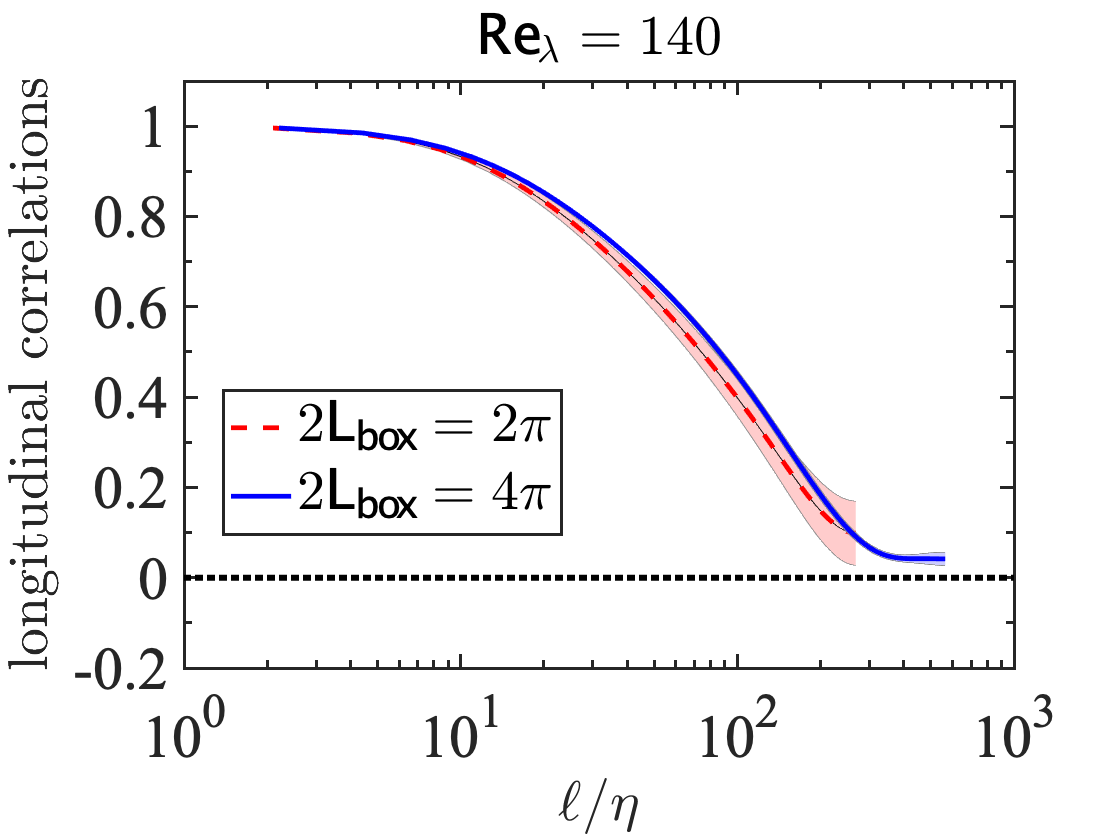}
\includegraphics[width=0.32\columnwidth]{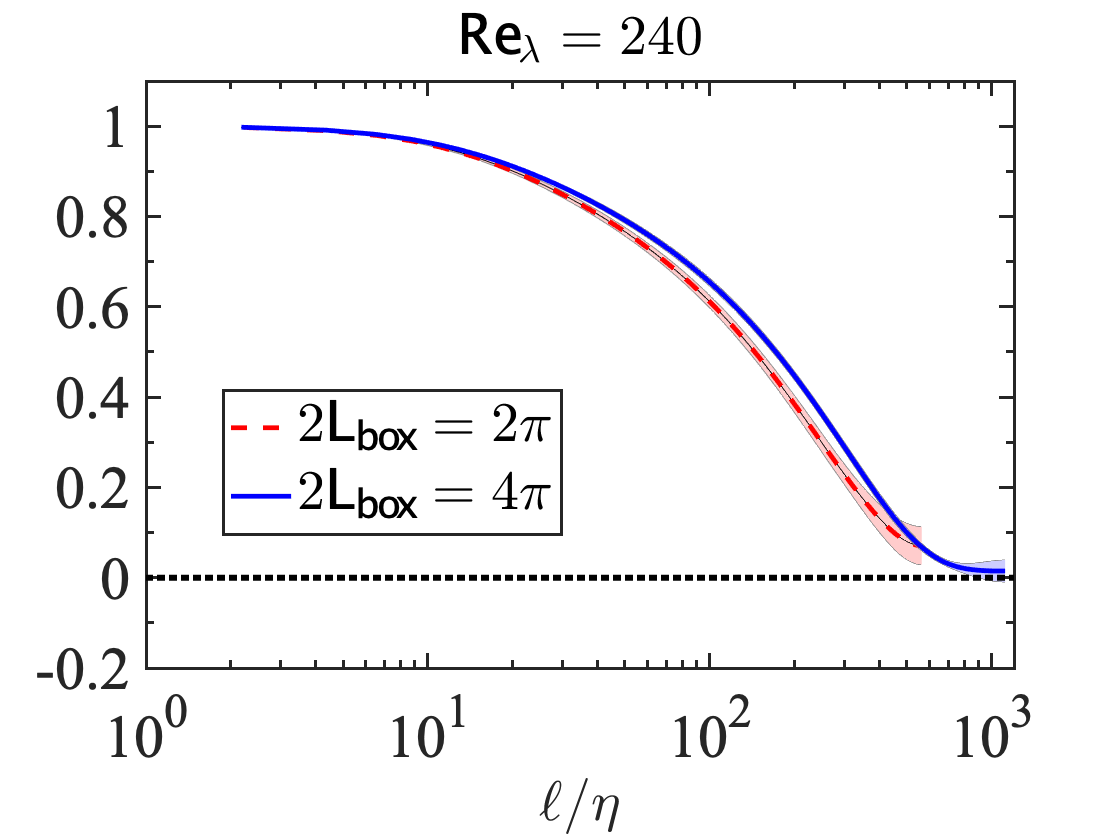}
\includegraphics[width=0.32\columnwidth]{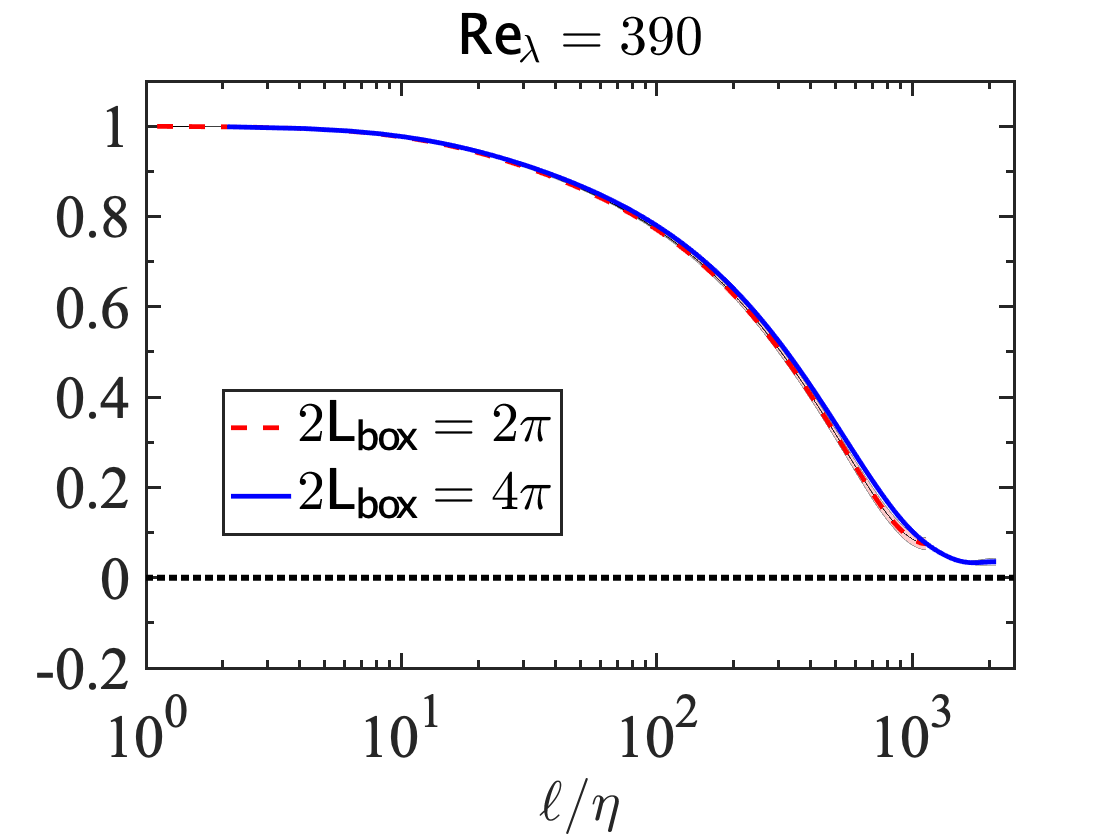} \\
\includegraphics[width=0.32\columnwidth]{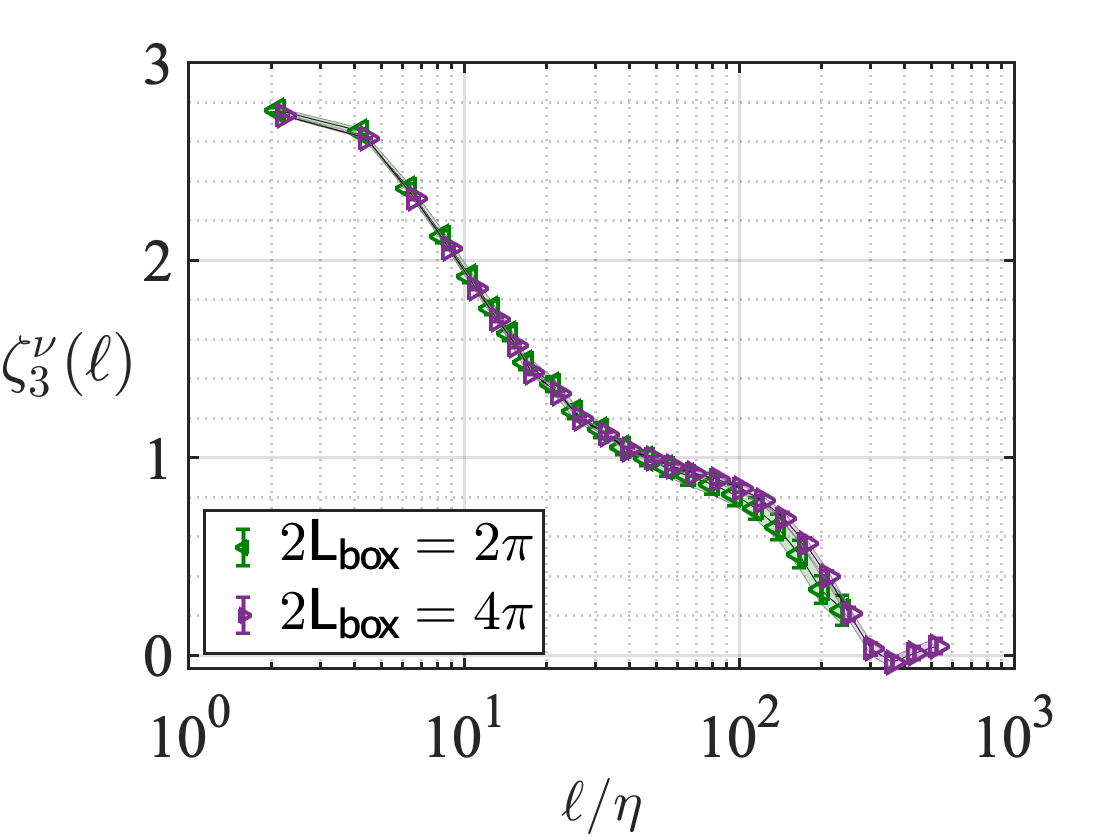}
\includegraphics[width=0.32\columnwidth]{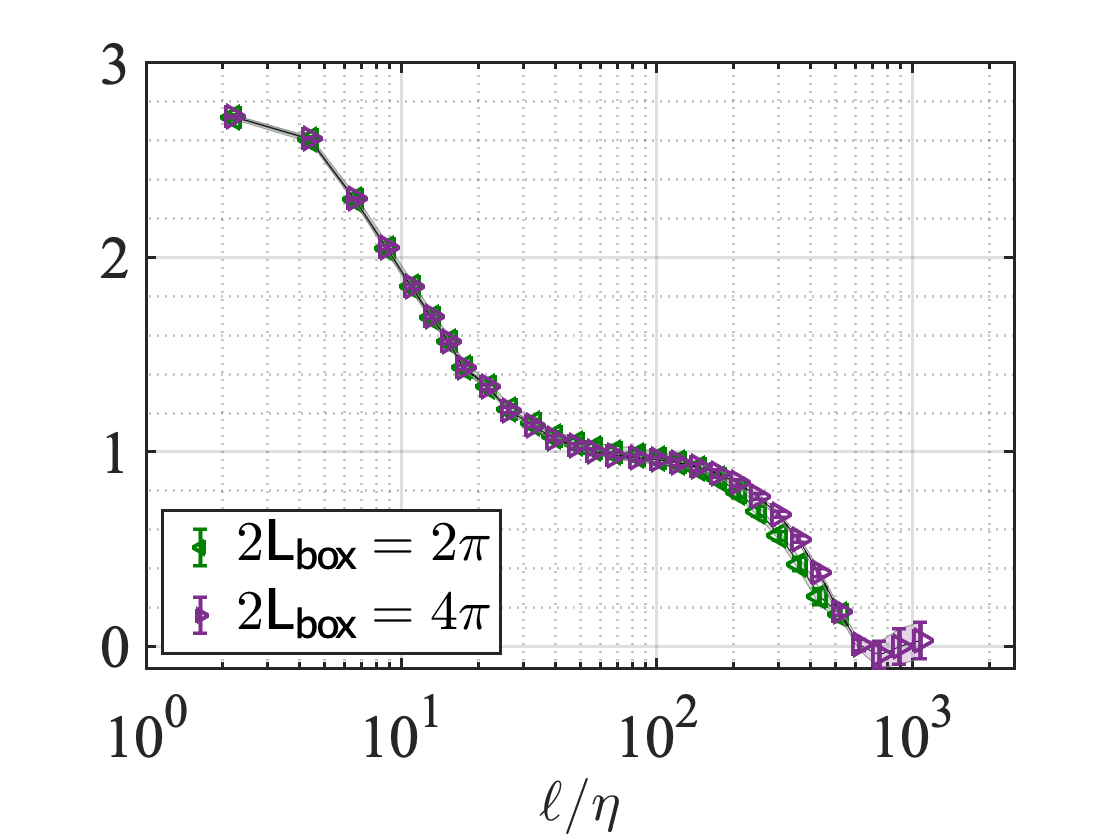}
\includegraphics[width=0.32\columnwidth]{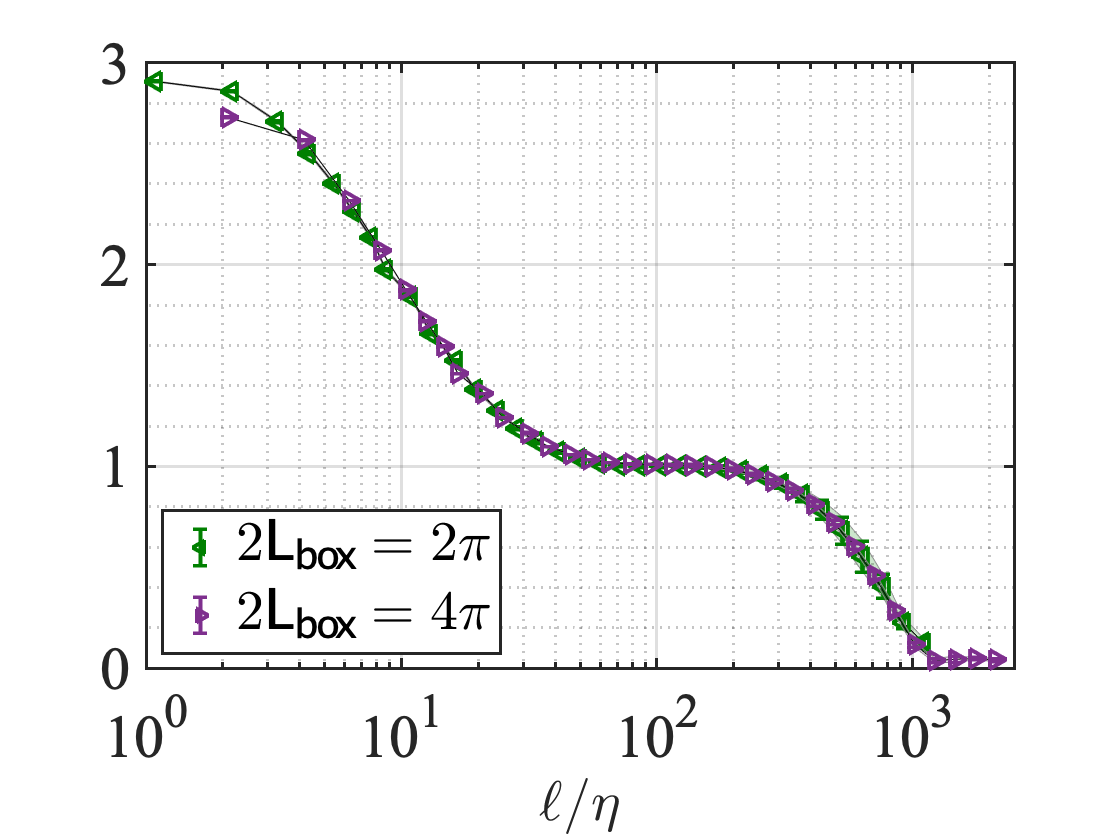}
\protect\caption{
Effects of box length and large-scale forcing in DNS of tri-periodic cubes with edge length $2 \lbox$. In each column DNS at two different edge lengths are shown with different forcing, but nominally the same $\rel$.
(Top row) One-dimensional two-point normalized longitudinal velocity correlations 
$\langle u(\ux) u(\ux+\ell \hat{\mathbf{e}}_x) \rangle/\langle u^2 \rangle$ where $u$ is the velocity along unit separation $\hat{\mathbf{e}}_x$
as a function of the scale-size $\ell$ normalized 
by the Kolmogorov scale $\eta$.
Shaded regions indicate the temporal standard deviations in the respective correlations.
Dotted line at zero is the expected large-$\ell$ estimate.
(Bottom row) third-order local slopes \eqref{lstot.eq} are
plotted in each panel for the two box lengths.
}
\label{boxlength.fig}
\end{figure}

\clearpage

\section*{Large scale properties} 
We briefly examine the dependence of the length and time scales that are relevant to the Reynolds number evolution of the normalized dissipation $\epsn \L/{\U}^3$ for the DNS data considered here: box length $2\lbox = 2\pi$ (length units) and frozen-energy forcing in the wavenumber range $0 < k \le k_F$, with $k_F = 2.1$ in all DNS runs. 
Figure \ref{Lint.fig} shows the dependence of the two choices of the large-length scale $\L$ considered in this work:
\beq 
\label{lscales.eq} \L=\lbox \;, \quad \L = \lint = \frac{\pi}{2{\up }^2}\int_{\klow}^{\kmax} \frac{E(k)}{k} dk \;, 
\eeq 
where $\up$ is the velocity rms, $\lbox$ is the half-length of the computational domain and $\lint$ is the isotropic integral scale obtained from the three-dimensional energy spectrum $E(k)$ at wavenumber $k$. Note that $\lint$ is usually adopted as the reference length in grid turbulence. 
The lower limit of the above integral $\klow$ is technically zero but in these simulations, it is truncated by the finite box size to $\pi/\lbox$. Likewise, the upper limit $\kmax$ is technically infinity, but is set to 
$(\sqrt{2}N/3) \klow$ for a DNS with $N^3$ collocation points.
This leads to a distorted estimate of $\lint$ (see earlier discussion on the persistent effects of large-scale anisotropy), but we shall nevertheless use it as a surrogate. Unlike the choice of the fixed half-box length of the DNS $\L = \lbox$, the integral scale computed as stated displays a slightly growing trend with $\re$, as shown in the inset. The $\re$-evolution of $\duint = \du (\lint/\lbox)$ shown in the main panel of Figure \ref{Lint.fig} is thereby consistent with that of $\du$ shown in the main text
and also with the upper bound in \cite{drivas2019onsager}, derived for a non-dimensionalization so that the box 
remains fixed as $\Re$ varies.  Note that the exponent arising from the use of $\lint$ is measurably smaller than that found by using $\lbox$. We also note that this growth trend cannot persist indefinitely since necessarily $\lint/\lbox\leq 2.$ Thus the discrepancy would eventually vanish.

\vspace{5pt} 

The evolution of the large-eddy time scale $T_E = \L/\up$  for the two choices of $\L$ \eqref{lscales.eq} are shown in Figure \ref{ts.fig} (left panel). Both estimates of $T_E$ appear to settle down nearly to a constant for $\re \gtrsim 10^4,$ the weak variation 
consistent with the inset in Figure \ref{Lint.fig}. On the other hand, the dissipation time-scale $T_\varepsilon = {\up}^2/\epsn$ shows a discernible increase with $\re,$ as seen in right panel of Figure \ref{ts.fig} in the same $\re$-range. 
This mismatch between the turbulent time-scale $T_E$ and the dissipation time-scale $T_\varepsilon$ results in their ratio $T_E/T_\varepsilon$ (which is the 
normalized dissipation $\epsn L/ {\up}^3$) to decay weakly with $\re$ at the rate quantified in the main text.

\begin{figure}[h!]
\centering
\includegraphics[width=0.5\columnwidth]{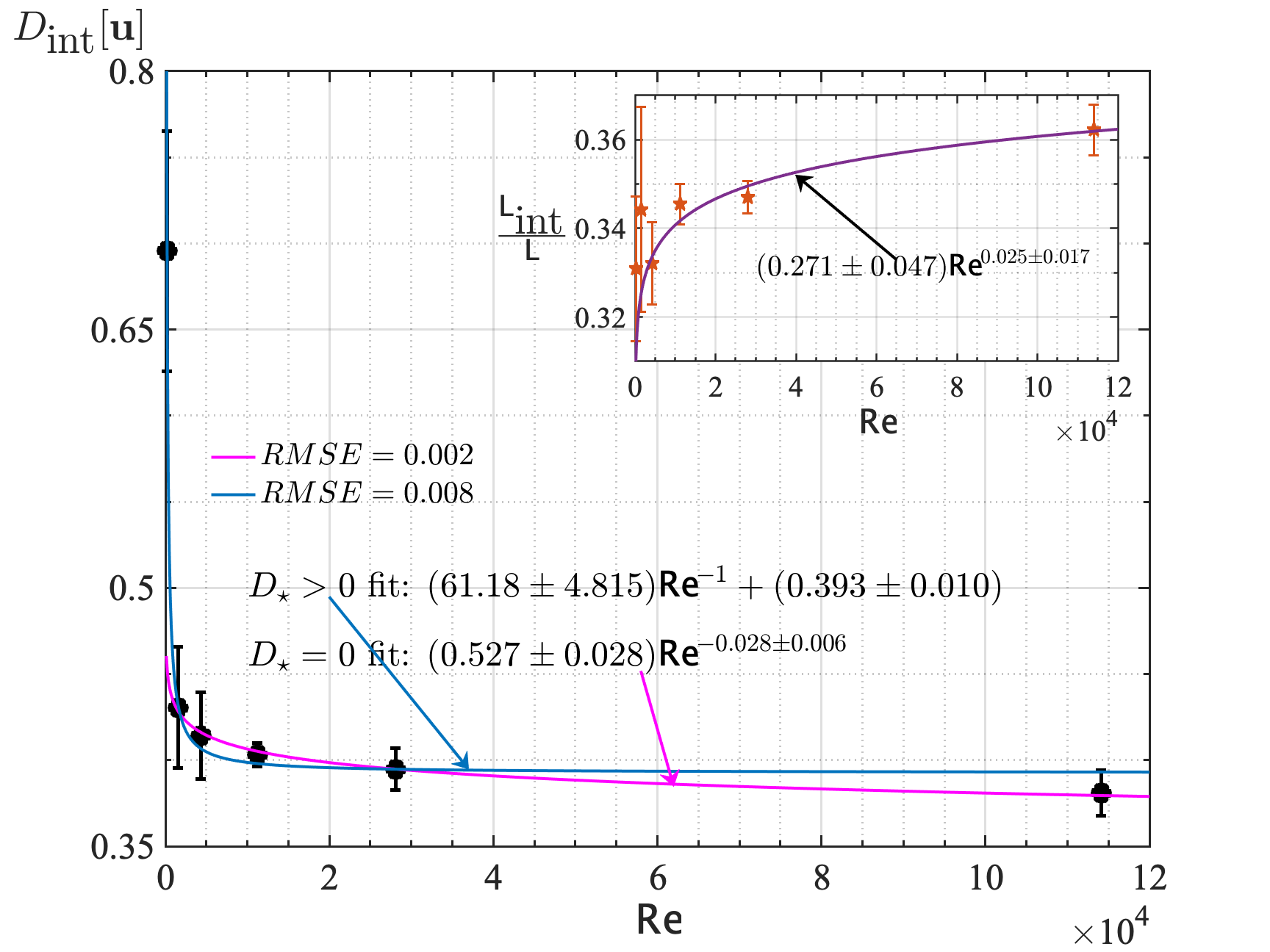}
\protect\caption{Same as Figure $2$ in main text but using $\lint$ \eqref{lscales.eq} as the length scale, $\duint = \epsn \lint/{\U}^3$. Fits corresponding to ($\dstar = 0$) $\zith > 1$ and ($\dstar > 0$) $\zith \le 1$ are given as are the respective RMSE. Inset shows weak growth of $\lint/\lbox$ when plotted against $\re$. 
}
\label{Lint.fig}
\end{figure}

\begin{figure}[h!]
\includegraphics[width=0.47\columnwidth]{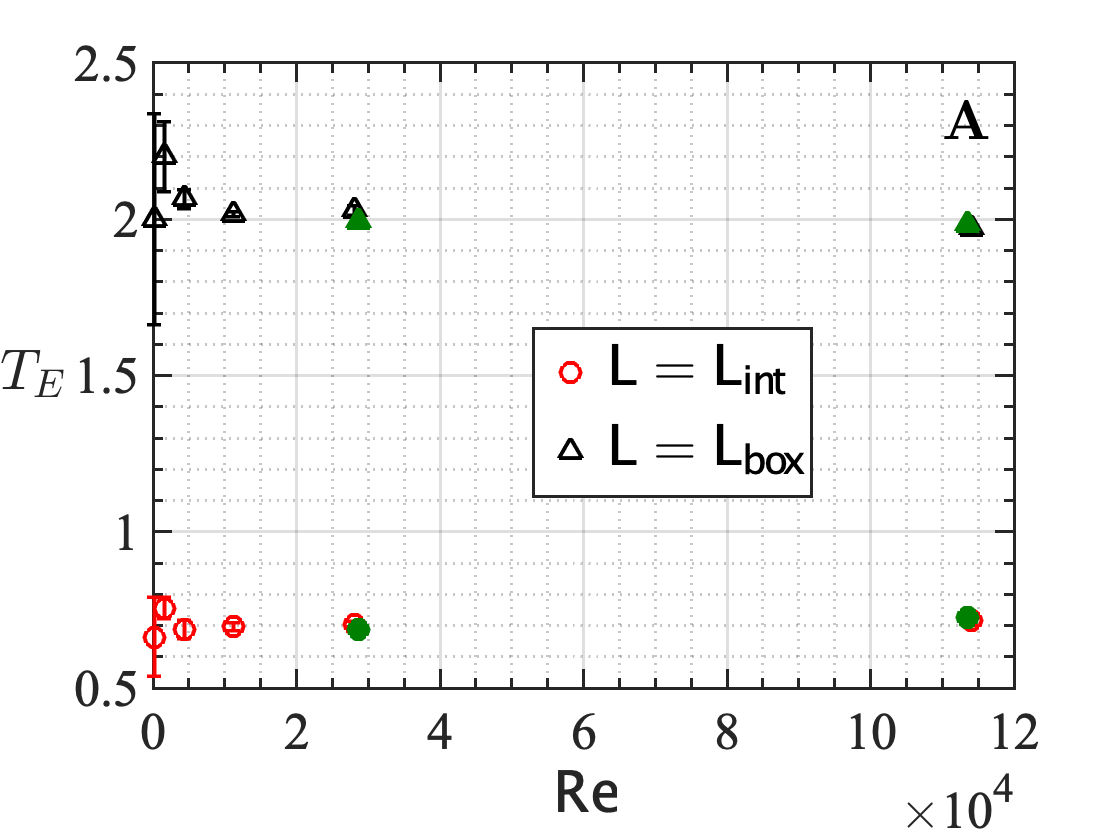}
\includegraphics[width=0.47\columnwidth]{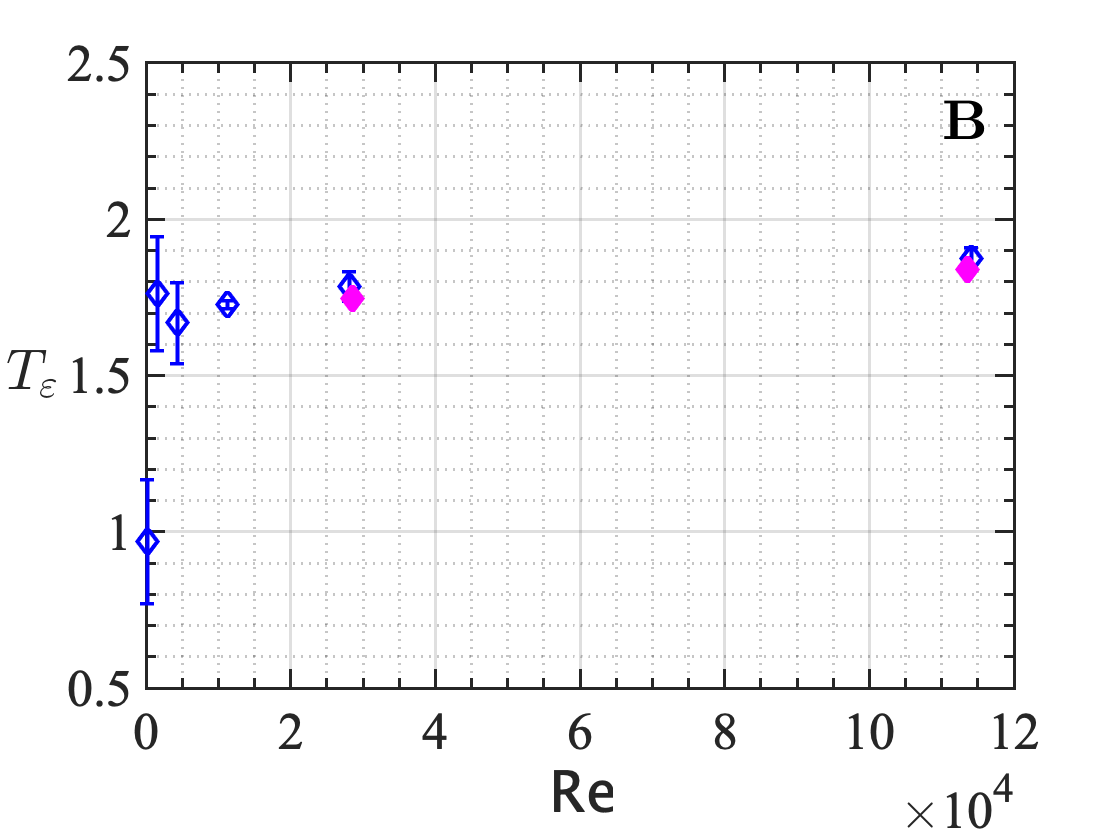} 
\protect\caption{
Turbulent time-scales as a function of Reynolds number $\re$ in DNS with side-length $2\lbox = 2\pi$.
(A) characteristic large-eddy time scale $T_E = \L/\up$ for 
two choices of length scale $\L$ used in this work.
(B) time scale of dissipation $T_\varepsilon = {\up}^2/\epsn$. Ordinate range in both panels is the same for comparison purposes. 
Filled symbols in both panels correspond to shorter DNS with higher spatial and temporal resolution \cite{iyer2020scaling}. 
}
\label{ts.fig}
\end{figure}

\clearpage 

\bibliography{bibliography}